\documentclass{aastex63}
\usepackage{xcolor}
\usepackage{longtable}
\usepackage{amsmath,amssymb}
\usepackage{rotating}

\usepackage{amssymb}
\usepackage{url}

\makeatletter

\newcommand{\Rmnum}[1]{\expandafter\@slowromancap\romannumeral #1@}
\makeatother

\submitjournal{ApJL}

\shorttitle{Insight-HXMT observed a giant outburst of 1A~0535+262}
\shortauthors{Kong et al.}

\begin{document}

\title{Luminosity dependence of the cyclotron line energy in 1A~0535+262 observed by Insight-HXMT during 2020 giant outburst}

\author{L. D. Kong\textsuperscript{*}}
\email{kongld@ihep.ac.cn}
\affil{Key Laboratory for Particle Astrophysics, Institute of High Energy Physics, Chinese Academy of Sciences, 19B Yuquan Road, Beijing 100049, China}
\affil{University of Chinese Academy of Sciences, Chinese Academy of Sciences, Beijing 100049, China}

\author{S. Zhang\textsuperscript{*}}
\email{szhang@ihep.ac.cn}
\affil{Key Laboratory for Particle Astrophysics, Institute of High Energy Physics, Chinese Academy of Sciences, 19B Yuquan Road, Beijing 100049, China}

\author{L. Ji\textsuperscript{*}}
\email{jilong@mail.sysu.edu.cn}
\affil{School of Physics and Astronomy, Sun Yat-Sen University, Zhuhai, 519082, China}

\author{P. Reig}
\affil{Institute of Astrophysics, Foundation for Research and Technology-Hellas, 71110 Heraklion, Crete, Greece}
\affil{University of Crete, Physics Department, 70013 Heraklion, Crete, Greece}

\author{V. Doroshenko}
\affil{Institut f{\"u}r Astronomie und Astrophysik, Kepler Center for Astro and Particle Physics, Eberhard Karls, Universit{\"a}t, Sand 1, D-72076 T{\"u}bingen, Germany}
\affil{Space Research Institute of the Russian Academy of Sciences, Profsoyuznaya Str. 84/32, Moscow 117997, Russia}

\author{A. Santangelo}
\affil{Institut f{\"u}r Astronomie und Astrophysik, Kepler Center for Astro and Particle Physics, Eberhard Karls, Universit{\"a}t, Sand 1, D-72076 T{\"u}bingen, Germany}

\author{R. Staubert}
\affil{Institut f{\"u}r Astronomie und Astrophysik, Kepler Center for Astro and Particle Physics, Eberhard Karls, Universit{\"a}t, Sand 1, D-72076 T{\"u}bingen, Germany}

\author{S. N. Zhang}
\affil{Key Laboratory for Particle Astrophysics, Institute of High Energy Physics, Chinese Academy of Sciences, 19B Yuquan Road, Beijing 100049, China}
\affil{University of Chinese Academy of Sciences, Chinese Academy of Sciences, Beijing 100049, China}

\author{R. Soria}
\affil{5College of Astronomy and Space Sciences, University of the Chinese Academy of Sciences, Beijing 100049, China}
\affil{Sydney Institute for Astronomy, School of Physics A28, The University of Sydney, Sydney, NSW 2006, Australia}

\author{Z. Chang}
\affil{Key Laboratory for Particle Astrophysics, Institute of High Energy Physics, Chinese Academy of Sciences, 19B Yuquan Road, Beijing 100049, China}

\author{Y. P. Chen}
\affil{Key Laboratory for Particle Astrophysics, Institute of High Energy Physics, Chinese Academy of Sciences, 19B Yuquan Road, Beijing 100049, China}

\author{P. J. Wang}
\affil{Key Laboratory for Particle Astrophysics, Institute of High Energy Physics, Chinese Academy of Sciences, 19B Yuquan Road, Beijing 100049, China}
\affil{University of Chinese Academy of Sciences, Chinese Academy of Sciences, Beijing 100049, China}

\author{L. Tao}
\affil{Key Laboratory for Particle Astrophysics, Institute of High Energy Physics, Chinese Academy of Sciences, 19B Yuquan Road, Beijing 100049, China}

\author{J. L. Qu}
\affil{Key Laboratory for Particle Astrophysics, Institute of High Energy Physics, Chinese Academy of Sciences, 19B Yuquan Road, Beijing 100049, China}
\affil{University of Chinese Academy of Sciences, Chinese Academy of Sciences, Beijing 100049, China}

\begin{abstract}
We report on a detailed spectral analysis of the transient X-ray pulsar 1A~0535+262, which underwent the brightest giant outburst ever recorded for this source from November to December 2020 with a peak luminosity of $1.2$ $\times10^{38}\ \rm erg\ s^{-1}$. 
Thanks to the unprecedented energy coverage and high cadence observations provided by Insight-HXMT, we were able to find for the first time evidence for a transition of the accretion regime.
At high luminosity, above the critical luminosity $6.7\times10^{37}$ erg s$^{-1}$, the cyclotron absorption line energy anti-correlates with luminosity.
Below the critical luminosity, a positive correlation is observed.
The 1A~0535+262 becomes, therefore, the second source after V~0332+53, which clearly shows an anti-correlation above and transition between correlation and anti-correlation around the critical luminosity.
The evolution of both the observed CRSF line energy and broadband X-ray continuum spectrum throughout the outburst exhibits significant differences during the rising and fading phases: that is,  for a similar luminosity the spectral parameters take different values which results in hysteresis patterns for several spectral parameters including the cyclotron line energy. 
We argue that, similarly to V~0332+53, these changes might be related to different geometry of the emission region in rising and declining parts of the outburst, probably due to a changes in the accretion disk structure and its interaction with the magnetosphere of the neutron star.

\end{abstract}
\keywords{pulsars: individual (A0535+26), X-rays: binaries, accretion: accretion pulsar}

\section{Introduction}
Accreting pulsars are neutron stars (NS) with a magnetic field of $B\sim$ $10^{12}$ G, orbiting and powered by accretion from a massive (spectral type late O or early B)  companion star (\citealp{Reig2013}). 
The strong magnetic field implies a relatively large margnetospheric radius of $\sim10^{7-10}$\,cm. 
At the magnetosphere the accreting plasma is forced to follow the magnetic field lines, and matter is channeled mostly onto the magnetic poles (mounds or columns depending on the luminosity).
In the hot, highly magnetized plasma the kinetic energy of the in-falling material is converted to heat and radiation. 
Typical cutoff power law continuum spectra and absorption features are generated in these structures. 

Magnetic field in many of these objects can be measured directly through measurement of energies of so-called cyclotron resonant scattering features (CRSFs) modifying the continuum X-ray spectrum of X-ray pulsars. 
These are characteristic absorption-like features found in the spectra of highly magnetized accreting pulsars usually from $\sim$ 10 keV to 100 keV corresponding to so-called Landau levels (\citealp{Staubert2019}) and directly related to field strength in the line forming region.  
The centroid line energies $E_{\rm cyc}$ are expected, i.e. at:
\begin{equation}
E_{cyc}\ =\ \frac{n}{(1+z)}\frac{\hbar e B}{m_{\rm e} c}\ \approx\  11.6\ \frac{n}{(1+z)} \times B_{12}\ (\rm keV),
\end{equation}
where $B_{12}$ is the magnetic field strength in units of $10^{12}$ G, $z$ is the gravitational redshift due to the NS mass and $n$ is the number of the Landau level involved: e.g., $n\ =\ 1$ for the fundamental line and $n\ \ge \ 2$ for the harmonics.

With the increase in precision of CRSF measurements, the dependence of CRSFs on luminosity and spin phase has become obvious and is one of the important areas of current X-ray pulsar research as it potentially allows probing of emission region geometry and properties (\citealp{Staubert2019}).
Of particular interest is luminosity dependence of line properties as it might be used to probe mechanism of plasma deceleration and transition between so-called sub- and super- critical accreition regimes associated with onset of an accretion column first suggested by \cite{Basko1976}.
As a result, two main accretion regimes, super-critical and sub-critical accretion, are expected depending on whether the source luminosity is higher or lower than a ``critical luminosity'' $L_{\rm crit}$, which strongly depends on the magnetic field of the neutron star (\citealp{Basko1976,Becker2012,Mushtukov2015MNRAS.447.1847M}). 
In the super-critical regime, for $L>L_{\rm crit}$, an accretion column forms, and the falling plasma is decelerated by a radiation shock that forms at a certain distance from the neutron star surface. 
In this case, the emission height increases with increasing luminosity. 
In the sub-critical regime, for  $L<L_{\rm crit}$, the in-falling matter is presumably decelerated by Coulomb interaction forming a region whose height decreases with increasing luminosity.
At even lower luminosity, the description of the deceleration process of falling material is not very conclusive. 
A reasonable scenario is that there exists a transition for the regime from Coulomb-stopping to gas-shock dominance beyond which only a small accretion mound forms on the NS surface.
Based on this picture, and considering that local magnetic field strength decreases with height, the correlation between the CRSF line energy and luminosity can be used to trace the accretion regimes. For instance, a transition between the sub and super-critical regimes has been recently reported by \cite{Doroshenko2017} and \cite{Vybornov2018} for V~0332+53 during a giant outburst.

So far a clear negative correlation between the $E_{\rm cyc}$ and the X-ray luminosity was only confirmed in V 0332+53 (\citealp{Makishima1990}, \citealp{Tsygankov2010}, \citealp{Cusumano2016}, \citealp{Doroshenko2017}; \citealp{Vybornov2018}), while for 4U 0115+63, the reported anti-correlation is debated (\citealp{Tsygankov2007}; \citealp{Nakajima2006}; \citealp{Mueller2013};
\citealp{Iyer2015}).
A positive correlation of the line centroid with luminosity is more common and it has been found in Her~X-1 (\citealp{Staubert2007}), Vela~X-1 (\citealp{Furst2014}),  GX~304-1 (\citealp{Klochkov2012}), and Cep~X-4 (\citealp{Furst2015}). 
For 1A~0535+262, the relationship between cyclotron energy and luminosity is not settled, with some results supporting a positive correlation (\citealp{Reig2013}, \citealp{Sartore2015}) and others reporting the absence of any correlation (\citealp{Terada2006}, \citealp{Caballero2007}; \citealp{Klochkov2011}). 
GX~304-1 and Vela X-1 also appear to show a flattening of the correlation with increasing luminosity (\citealp{Rothschild2017}, \citealp{Parola2016}).
Here we investigate luminosity dependence of line properties in another Be XRB transient 1A~0535+262 at luminosities exceeding those reached in previous investigations. 

The transient X-ray binary 1A~0535+262 was discovered in outburst during observations of the Crab by the Rotation Modulation Collimator on $Ariel\ V$, through detection of the pulsations with period of $\sim$ 104\,s (\citealp{Rosenberg1975}). 
The system is composed of an O9.7IIIe donor star and a strongly magnetized neutron star (\citealp{Steele1998}). 
The orbit has an eccentricity of $e=0.47$ and an orbital period of $P_{\rm orb}\sim110.3$\,d (\citealp{Finger1996}). 
The distance to 1A~0535+262 is 2\,kpc, which has been recently confirmed by Gaia (\citealp{Bailer-Jones2018}).
The fundamental line and the first harmonic have been detected at $\sim$ 46\,keV and $\sim$ 100\,keV by different missions during different outbursts (\citealp{Kendziorra1994}; \citealp{Grove1995}; \citealp{Staubert2019}). 
A somewhat higher CRSF line energy of $\sim$ 50 keV was found during the pre-outburst flare which took place about 5 days before the peak of a normal (type-I) outburst (\citealp{Caballero2008}; \citealp{Postnov2008}). 

Since its discovery, 1A~0535+262 experienced a series of outbursts with peak flux varying from a few 100 mCrab to $\sim$ 5.5 Crab (\citealp{Camero-Arranz2012}, \citealp{Caballero2013}). 
After a long period of quiescence following the 2011 outburst, the source went through a giant outburst in November 2020, with a peak flux recorded as $\sim$ 12.5 Crab in the energy range $15-50$\,keV by Swift/BAT.
Owing to the unprecedentedly high intrinsic source flux, observation cadence and wide energy range covered by Insight-HXMT, we can investigate for the first time:

\begin{itemize}
\item The relationship between the CRSF energy and luminosity at a higher luminosity than ever before

\item The evolution of the spectral parameters, both during the rise and decline of the outburst, allowing us to study whether the evolutionary paths are similar or differ (hysteresis) at different phases of the outburst.
\end{itemize}
Sec~2 briefly describes the Insight-HXMT detectors and the data reduction procedure.
The details of the spectral analysis are presented in Sec~3.1, while Sec~3.2 \& Sec~3.3 focus on the evolution of the CRSF with luminosity and time, respectively. 
Sec~3.4 focus on the variation of other spectral components. 
Discussion and conclusions are included in Sec~4 and Sec~5.

\section{Observation and Data reduction}
\begin{figure}
    \centering\includegraphics[scale=0.5]{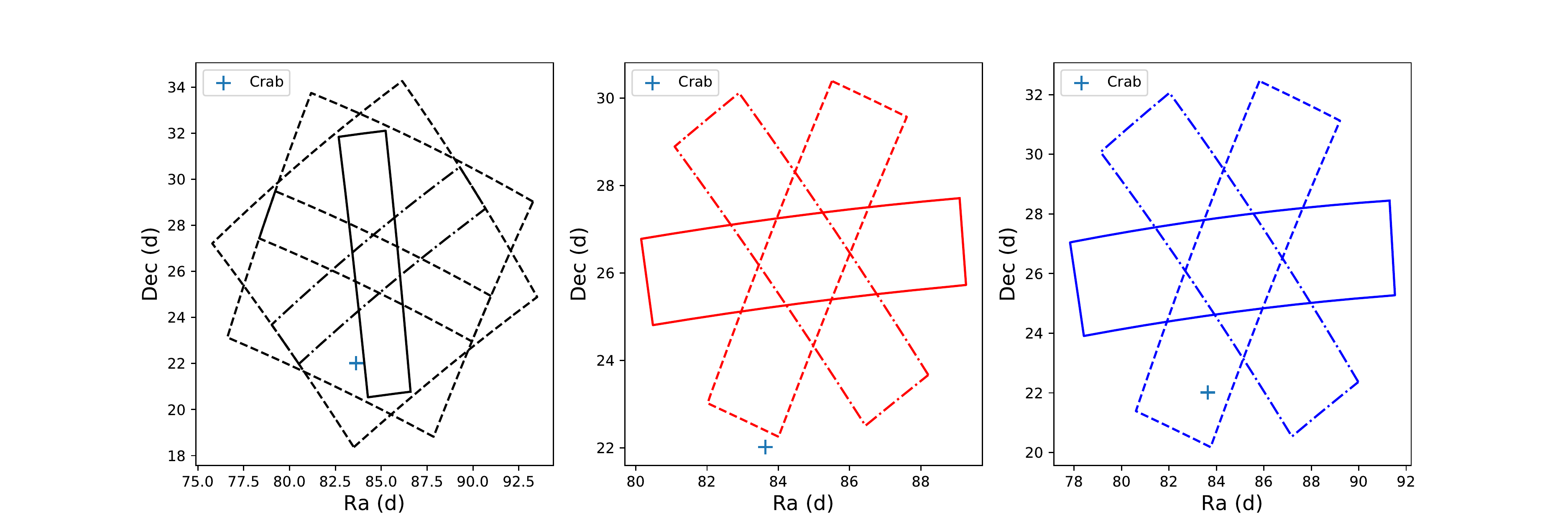}
    \caption{The FoVs for each detector are shown in black color for HE (three boxes for small FoVs and two boxes for large FoVs), red color for ME and blue color for LE, respectively. The blue cross shows the Crab location during the ObsID P031431600101.
    }
    \label{the FoVS}
\end{figure}

The Hard X-ray Modulation Telescope, also dubbed as Insight-HXMT (\citealp{2014SPIE.9144E..21Z}; \citealp{2020SCPMA..63x9502Z}), was launched on June 15, 2017, with a science payload that allows observations in a broad energy band (1-250 keV) and with large effective area at high energies.
Insight-HXMT consists of three collimated telescopes: the High Energy X-ray Telescope (HE, 18 cylindrical NaI(Tl)/CsI(Na) phoswich detectors) (\citealp{2020SCPMA..63x9503L}), the Medium Energy X-ray Telescope (ME, 1728 Si-PIN detectors) (\citealp{2020SCPMA..63x9504C}), and the Low Energy X-ray Telescope (LE, Swept Charge Device (SCD)) (\citealp{2020SCPMA..63x9505C}), with collecting-area/energy-range of 5000 $\rm cm^2$/20-250 keV, 952 $\rm cm^2$/5-30 keV and 384 $\rm cm^2$/1-10 keV, and typical Field of View (FoV) of $1.6^{\circ}\times6^{\circ}$, $1^{\circ}\times4^{\circ}$ and $1.1^{\circ}\times5.7^{\circ}$; $5.7^{\circ}\times5.7^{\circ}$ for LE, ME and HE, respectively (see Figure~\ref{the FoVS}). 
Insight-HXMT observed 1A~0535+262 from Nov 6, 2020 (MJD 59159) to Dec 24, 2020 (MJD 59207), for a total exposure of $\sim$ 1.910\,Ms.
For the analysis reported in this article, we used the Insight-HXMT Data Analysis Software (HXMTDAS) v2.04, together with the current calibration model v2.05 (\url{http://hxmtweb.ihep.ac.cn/software.jhtml}). 
The data are selected under a series of criteria as recommended by the Insight-HXMT team. In particular, data with elevation angle (ELV) larger than $10^{\circ}$, geometric cutoff rigidity (COR) larger than 8 GeV and offset for the point position smaller than $0.04^{\circ}$ were considered.
In addition data taken within 300\,s of the South Atlantic Anomaly (SAA) passage outside of good time intervals identified by onboard software was rejected. 
Because the Crab (ra=83.63308, dec=22.0145) is located close to 1A~0535+262 (ra=84.727396, dec=26.315789), and occasionally falls into the FOV of some of the Insight-HXMT detectors, we exclude also the detector boxes which suffered from contamination by the Crab (see Figure~\ref{the FoVS}).

The energy bands adopted for the spectral analysis are: $2-7$\,keV for the LE, $8-35$\,keV for the ME, and $30-120$\,keV for the HE. 
Because of the uncertain calibration at 22 keV of the ME, we ignore the $20-23$ keV range during the spectral analysis for this instrument. 
The instrumental backgrounds are estimated with the tools provided by Insight-HXMT team: LEBKGMAP, MEBKGMAP and HEBKGMAP, version 2.0.9 based on the standard Insight-HXMT background models (\citealp{2020arXiv200401432L}; \citealp{2020arXiv200501661L}; \citealp{2020arXiv200306260G}). 
The XSPEC v12.10.1f software package (\citealp{1996ASPC..101...17A}) was used to perform spectral fitting. In order to improve the counting statistic of the energy spectra, we combined the exposures within one day by \emph{addspec} and \emph{addrmf} tasks.
Considering current accuracy of the instrument calibration, we include 0.5$\%$, 0.5$\%$ and 1\% systematic error for spectral analysis for LE, ME and HE, respectively.
The uncertainties of the spectral parameters are computed using Markov Chain Monte Carlo (MCMC) with a length 10,000 and are reported at 90\% confidence level.

\section{Results}
\subsection{The spectral analysis}
\begin{figure}
    \centering\includegraphics[angle=-90,width=0.49\textwidth]{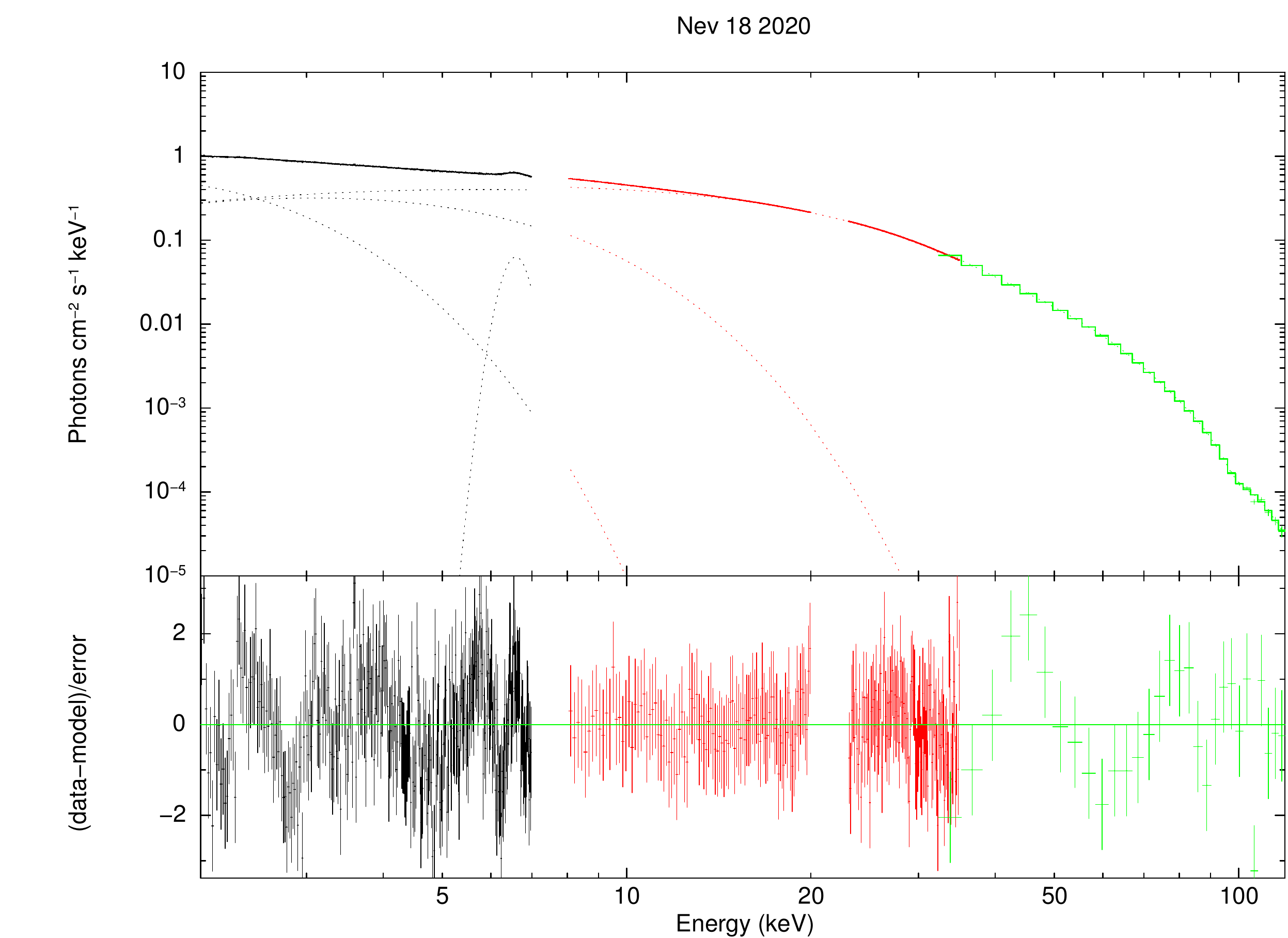}
    \centering\includegraphics[angle=-90,width=0.49\textwidth]{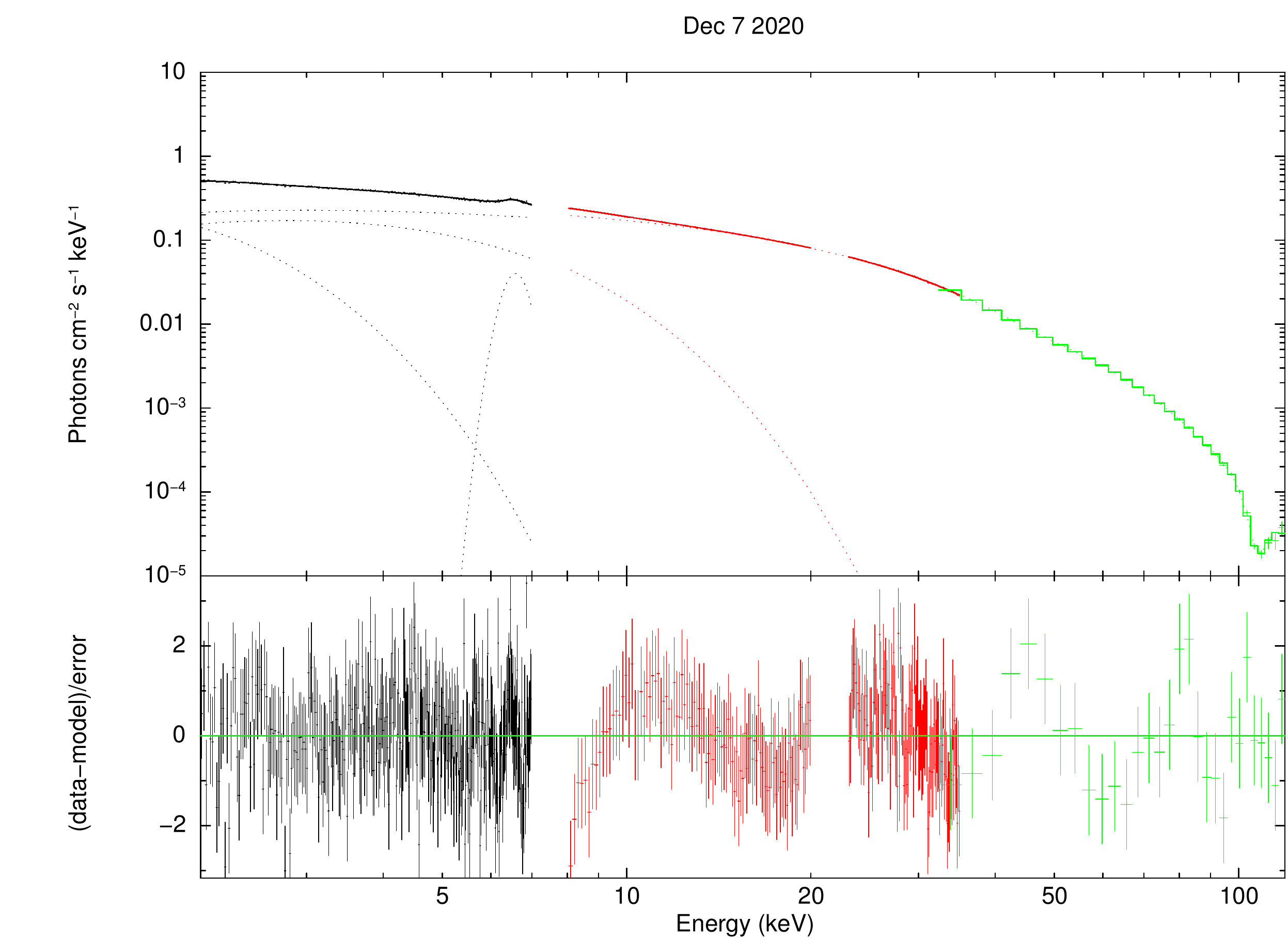}
    \centering\includegraphics[angle=-90,width=0.49\textwidth]{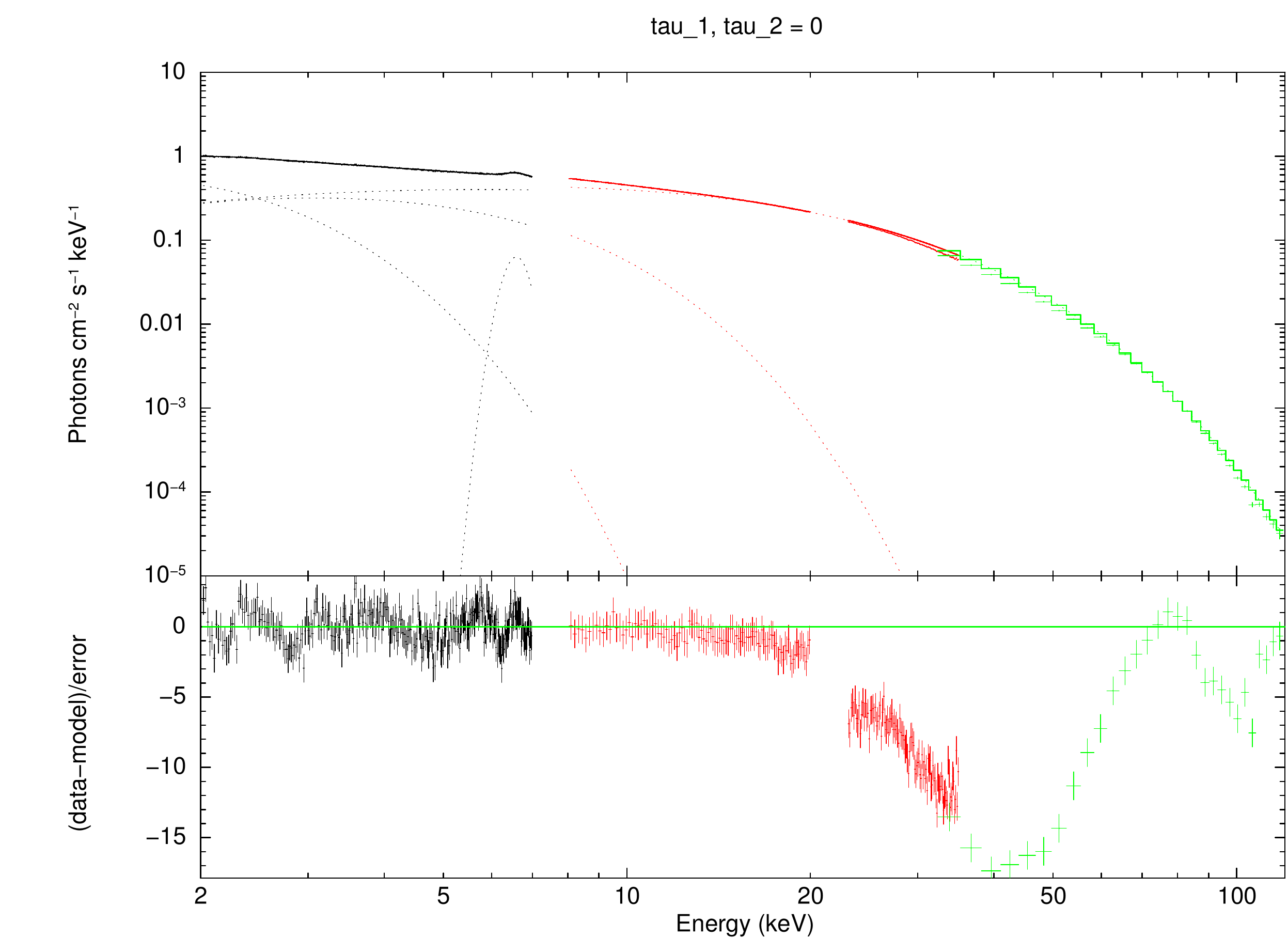}
    \centering\includegraphics[angle=-90,width=0.49\textwidth]{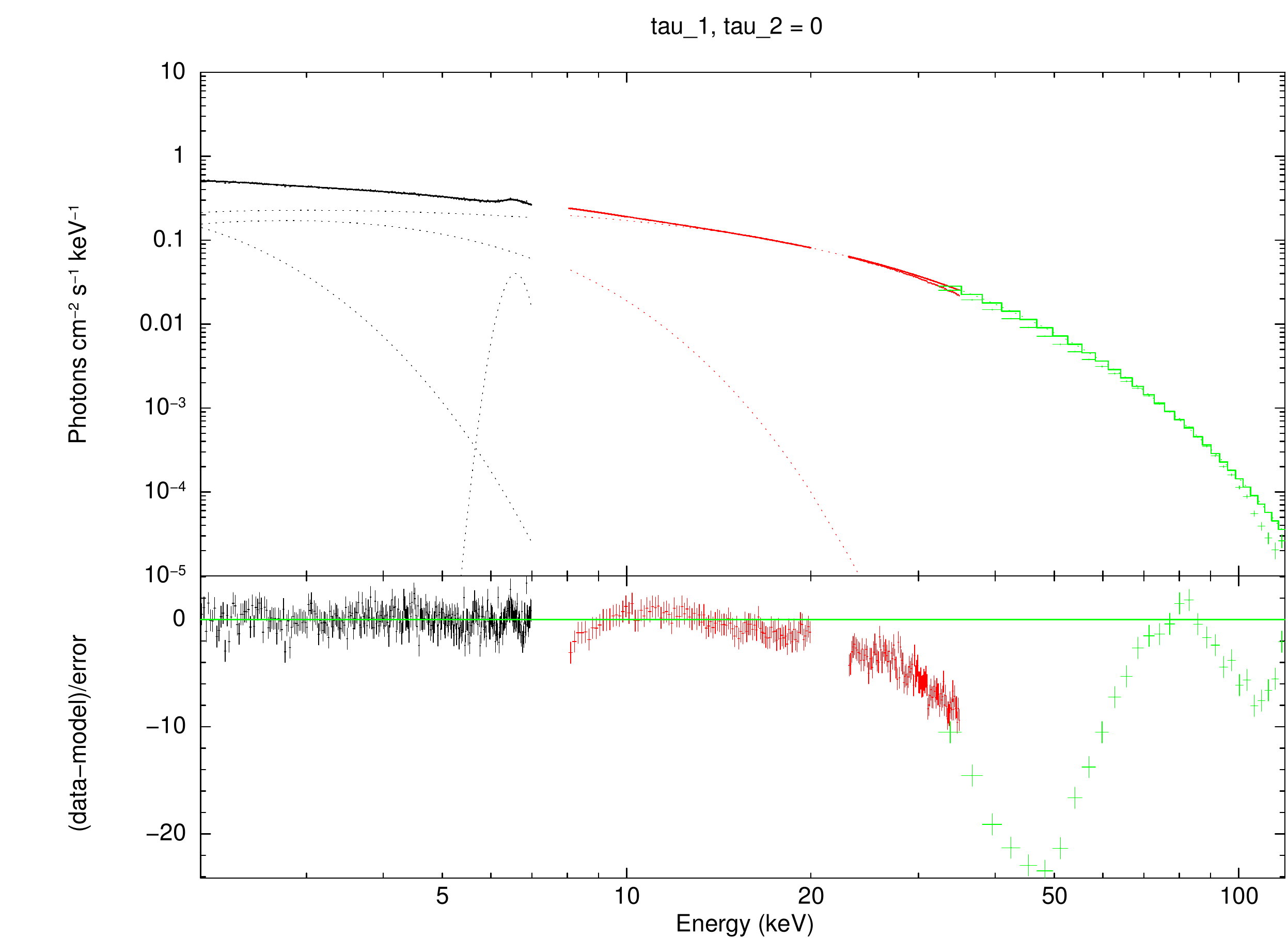}
    \caption{The panels show the spectra and residuals of Nov 18, 2020 and Dec 7, 2020. The spectra and residuals from LE, ME and HE detectors are shown in black, red and green points. Two black-body, a cutoffPL and a iron emission line are shown in the panels. The residuals without CRSF lines ($\tau_1, \tau_2 = 0$) are shown at the bottom of the figure.
    The parameters are shown in Table~\ref{spectral_fitting}.}
    \label{specta_plt}
\end{figure}

Considering the large number of observations carried out, we only show detailed spectral fits for two representative observations carried out on Nov 18, 2020 and Dec 7, 2020, respectively. 
The observation of Nov 18, 2020 (MJD 59171) corresponds to the peak of the outburst at a luminosity of $11.67 \times 10^{37}$ erg s$^{-1}$, while the X-ray luminosity during the observation on Dec 7, 2020 (MJD 59191) was $4.74 \times 10^{37}$ erg s$^{-1}$ (here and throughout the manuscript the luminosity is calculated in $2-150$ keV assuming a distance of 2~kpc (\citealp{Bailer-Jones2018})).  
In Figure~\ref{specta_plt}, the data from LE, ME, and HE are plotted in black, red and green colors, respectively. 
We first fitted the spectra with the following model: \emph{tbabs}$\times$(\emph{cutoffpl}+\emph{gaussian}). 
The \emph{cutoffpl} is a simple continuum with just three free parameters:
\begin{equation}
F(E)=K \times E^{-\Gamma}exp(-E/E_{\rm fold}) 
\end{equation}
where $K$, $\Gamma$ and $E_{\rm fold}$ determine the normalization coefficient, the photon index, and the exponential folding energy, respectively.
Absorption is taken into account through the \emph{tbabs} model, the Tuebingen-Boulder interstellar medium (ISM) absorption model (\citealp{Wilms2000}). 
The Galactic hydrogen column density toward the pulsar,  $n_{\rm H}$, is fixed at $0.59 \times 10^{22}$ atom cm$^{-2}$. 
The \emph{gaussian} line is needed to model the iron emission line, usually at $E\sim6.6$\,keV.

Significant residuals at $2-20$\,keV, $\sim 45$\,keV and $\sim100$\,keV are observed.
The residuals at lower energies were accounted for by adding two black-body components: a hotter one with $kT>1$ keV and a cooler one with $kT<1$ keV. 
The flux in $2-10$ keV of two black body and cutoffpl components are shown in Table~\ref{spectral_fitting}. 
Within $2-10$ keV, the flux of two black body component and that of cutoffpl component were comparable.
The absorption features are clearly visible in the residuals of best-fit spectra with the continuum model described above at $\sim45$ and $\sim 100$\,keV which is consistent with the previously determined energies of the fundamental CRSF and and its harmonic in 1A~0535+262 (\citealp{Kendziorra1994,Grove1995,Staubert2019}), and thus were interpreted as such. 
To model these lines, we used a multiplicative absorption model \emph{mgabs} with Gaussian profile:
\begin{equation}
F^{'}(E) = F(E) \times \emph{mgabs} = F(E) \times [(1-\tau_{1}e^{\frac{-(E-Ecyc)^2}{2\sigma_{1}^{2}}}) \times (1-\tau_{2}e^{\frac{-(E-R \times Ecyc)^2}{2\sigma_{2}^{2}}})], 
\end{equation}
where $F^{'}(E)$ is the spectrum modified by \emph{mgabs}, $E_{\rm cyc}$ is the cyclotron line central energy, $\tau_{1}$ and $\sigma_{1}$ characterize the central optical depth and the width of the line. 
The parameters $R$ describe the ratio of the 1-st harmonic line and fundamental line, $\tau_{2}$ and $\sigma_{2}$ describe the central optical depth and the width of the 1-st harmonic line.

The model \emph{tbabs} $\times$ \emph{magbs} $\times$ (\emph{bbodyrad1}+\emph{bbodyrad2}+\emph{gaussian}+\emph{cutoffpl}) improved the fit considerably giving a reduced $\chi^{2}$ of 0.97 and 0.93 for the two observations on Nov 18 and Dec 7, 2020, respectively (Table~\ref{spectral_fitting}). Although some residuals are still present at around 60 keV (Figure~\ref{specta_plt}), these are not significant (F-value = 4.5 \& 4.2, see Table~\ref{spectral_fitting}) and other parameters are not affected by the addition of an extra absorption component.
This structure may be caused by the non-Gaussian character of the cyclotron line or the uncertainty of the calibration. Because the significance of this structure is not high, it is difficult to draw any firm conclusion.

The fundamental $E_{\rm cyc}$ was 43\,keV during the peak of the outburst and increased to 47 keV during the fading phase of the outburst.
The ratio between the fundamental and harmonic line is $R \approx$ 2.3, and this ratio is also consistent with the results from INTEGRAL observations (\citealp{Sartore2015}).
The width of the fundamental is about half of
that of the harmonic, again consistent with previous observations by other satellites, where the width for fundamental line was measured to be $\approx $ 10 keV, and $\approx$ 5 keV for the harmonic (\citealp{Sartore2015}). 

The parameters of other spectral components also changed significantly between these two observations.
For the observation at higher luminosity, the temperatures of the black body components are $0.6_{-0.01}^{+0.01}$ keV and $1.9_{-0.02}^{+0.03}$ keV, higher than those ($0.47_{-0.01}^{+0.01}$\,keV and $1.55_{-0.01}^{+0.03}$\,keV) of the observation at lower luminosity. 
The normalization factors of the black body components between the two observations vary from $3999_{-224}^{+89}$ and $166_{-10}^{+6}$ to $3151_{-300}^{+363}$ and $115_{-6}^{+4}$.

To study the evolution of the fundamental CRSF line, we initially let all spectral parameters free, except the hydrogen column density that was fixed to $5.9\times 10^{21}$ cm$^{-2}$. 
Then we also tried fixing $R$, $\sigma_{1}$, $\sigma_{2}$, $\sigma_{\rm Fe}$ to the average values 2.3, 10 keV, 5 keV and 0.3 keV, respectively. 
Whether these parameters were fixed or not made no significant differences to the other parameters.
We will show the details of the spectral evolution in the following sections.

\begin{table}[ptbptbptb]
    \begin{center}
\caption{Parameters of the spectral fitting}
    \begin{tabular}{cccccccccccc}
\hline
\hline
Date & & \multicolumn{3}{c}{2020-11-18} & \multicolumn{3}{c}{2020-12-07}
\\
\hline
MJD & & \multicolumn{3}{c}{59171} & \multicolumn{3}{c}{59190}
\\
\hline
Model & Parameters & \multicolumn{3}{c}{2bb+cutoffPL} &  \multicolumn{3}{c}{2bb+cutoffPL}
\\
\hline
tbabs & $n_{\rm H}\ (10^{22}\ \rm cm^{-2})$ & $0.59$ (fixed) & $0.59$ (fixed) & $0.59$ (fixed) & $0.59$ (fixed) & $0.59$ (fixed) & $0.59$ (fixed)
\\
gabs & $E_{\rm abs}$ (keV) & ... & ... & $59.9_{-3.9}^{+7.2}$ & ... & ... & $61.0_{-2.5}^{+4.8}$ &
\\
& $\sigma_{\rm abs}$ (keV) & ... & ... & $3.8_{-3.3}^{+2.1}$ & ... & ... & $3.3_{-2.1}^{+5.6}$ &
\\
& $\tau_{\rm abs}$ & ... & ... & $0.06_{-0.03}^{+0.03}$ & ... & ... & $0.07_{-0.05}^{+0.04}$ &
\\
F-test & F-value & ... & ... & 4.5 & ... & ... & 4.2
\\
& P-value & ... & ... & 0.0035 & ... & ... & 0.0063
\\
mgabs & $E_{\rm cyc}$ (keV) & $43.0_{-0.4}^{+0.8}$ & $43.0_{-0.2}^{+0.4}$ & $42.4_{-0.3}^{+0.8}$ & $47.0_{-0.2}^{+1.0}$ & $46.7_{-0.3}^{+0.2}$ & $46.2_{-0.4}^{+0.6}$
\\
& $\sigma_{1}$ (keV) & $13_{-1}^{+1}$ & 10 (fixed) & 10 (fixed) & $11_{-1}^{+1}$ & 10 (fixed) & 10 (fixed)
\\
& $\tau_{1}$ & $0.24_{-0.01}^{+0.01}$ & $0.18_{-0.01}^{+0.01}$ & $0.18_{-0.01}^{+0.01}$ & $0.24_{-0.01}^{+0.01}$ & $0.24_{-0.01}^{+0.01}$ & $0.24_{-0.01}^{+0.01}$
\\
& R & $2.3_{-0.1}^{+0.1}$ & 2.3 (fixed) & 2.3 (fixed) & $2.3_{-0.1}^{+0.1}$ & 2.3 (fixed) & 2.3 (fixed)
\\
& $\sigma_{2}$ (keV) & $5_{-2}^{+3}$ & 5 (fixed) & 5 (fixed) & $8_{-2}^{+7}$ & 5 (fixed) & 5 (fixed) 
\\
& $\tau_{2}$ & $0.3_{-0.1}^{+0.1}$ & $0.3_{-0.1}^{+0.1}$ & $0.32_{-0.02}^{+0.04}$ & $0.6_{-0.1}^{+0.1}$ & $0.8_{-0.1}^{+0.1}$ & $0.8_{-0.1}^{+0.2}$
\\
gaussian & $E_{\rm Fe}$ (keV) & $6.58_{-0.02}^{+0.01}$ & $6.59_{-0.02}^{+0.02}$ & $6.59_{-0.01}^{+0.02}$ & $6.57_{-0.02}^{+0.01}$ & $6.58_{-0.02}^{+0.02}$ & $6.59_{-0.01}^{+0.01}$
\\
& $\sigma_{\rm Fe}$ (keV) & $0.19_{-0.03}^{+0.02}$ & 0.3 (fixed) & 0.3 (fixed) & $0.25_{-0.03}^{+0.03}$ & 0.3 (fixed) & 0.3 (fixed)
\\
& norm ($10^{-2}$)& $3.3_{-0.3}^{+0.1}$ & $4.8_{-0.3}^{+0.2}$ & $4.7_{-0.4}^{+0.3} $ & $2.6_{-0.3}^{+0.3}$ & $3.0_{-0.2}^{+0.1}$ & $3.0_{-0.1}^{+0.3}$
\\
bbodyrad2 & $kT$ (keV) & $0.60_{-0.01}^{+0.01}$ & $0.56_{-0.01}^{+0.01}$ & $0.57_{-0.01}^{+0.01}$ & $0.47_{-0.01}^{+0.01}$ & $0.44_{-0.02}^{+0.02}$ & $0.45_{-0.02}^{+0.01}$
\\
& norm & $3999_{-224}^{+89}$ & $4439_{-284}^{+276}$ & $4348_{-171}^{+294}$ & $3151_{-300}^{+363}$ & $3799_{-681}^{+666}$ & $3625_{-173}^{+772}$
\\
& $log_{10}$(Flux) & ... & $-8.63_{-0.02}^{+0.03}$ & ... & ... & $-9.31_{-0.04}^{+0.09}$ & ...
\\
& ($2-10$ keV, erg cm$^{-2}$ s$^{-1}$) & & & & & &
\\
bbodyrad1 & $kT$ (keV) & $1.90_{-0.02}^{+0.03}$ & $1.70_{-0.03}^{+0.03}$ & $1.73_{-0.03}^{+0.03}$ & $1.55_{-0.01}^{+0.03}$ & $1.50_{-0.02}^{+0.02}$ & $1.52_{-0.03}^{+0.01}$
\\
& norm & $166_{-10}^{+6}$ & $175_{-8}^{+9}$ & $173_{-8}^{+8}$ & $115_{-6}^{+4}$ & $123_{-8}^{+5}$ & $122_{-3}^{+8}$
\\
& $log_{10}$(Flux) & ... & $-7.90_{-0.02}^{+0.02}$ & ... & ... & $-8.24_{-0.02}^{+0.02}$ & ...
\\
& ($2-10$ keV, erg cm$^{-2}$ s$^{-1}$) & & & & & &
\\
cutoffPL & $\Gamma$ & $-0.80_{-0.02}^{+0.02}$ & $-0.55_{-0.02}^{+0.01}$ & $-0.57_{-0.02}^{+0.02}$ & $-0.10_{-0.02}^{+0.02}$ & $-0.08_{-0.02}^{+0.01}$ & $-0.09_{-0.01}^{+0.02}$
\\
& $E_{\rm fold}$ & $9.45_{-0.04}^{+0.07}$ & $10.04_{-0.06}^{+0.05}$ & $10.01_{-0.06}^{+0.05}$ & $12.3_{-0.1}^{+0.2}$ & $12.4_{-0.1}^{+0.1}$ & $12.4_{-0.1}^{+0.1}$
\\
& $K$ & $0.16_{-0.01}^{+0.01}$ & $0.28_{-0.01}^{+0.01}$ & $0.26_{-0.01}^{+0.01}$ & $0.27_{-0.01}^{+0.02}$ & $0.28_{-0.01}^{+0.01}$ & $0.28_{-0.01}^{+0.02}$
\\
& $log_{10}$(Flux) & ... & $-7.53_{-0.01}^{+0.01}$ & ... & ... & $-7.84_{-0.01}^{+0.01}$ & ...
\\
& ($2-10$ keV, erg cm$^{-2}$ s$^{-1}$) & & & & & &
\\
Luminosity & $L_{2-150}/10^{37}$ (erg s$^{-1}$)& ... & $11.67_{-0.05}^{+0.03}$ & ... & ... & $4.74_{-0.02}^{+0.01}$ & ...
\\
Fitting & $\chi_{\rm red}^{2}$/d.o.f & 0.97/510 & 1.03/514 & 1.01/511 & 0.93/510 & 0.95/514 & 0.93/511
\\
\hline
\hline
    \end{tabular}
    \label{spectral_fitting}
\begin{list}{}{}
    \item[Note]{: Uncertainties are reported at the 90\% confidence interval and were computed using MCMC (Markov Chain Monte Carlo) of length 10,000. The 0.5\%, 0.5\% and 1\% system error for LE, ME and HE respectively has been added during spectral fittings.}
\end{list}
    \end{center}
\end{table}

\subsection{The CRSF centroid energy evolution}
\begin{figure}
    \centering\includegraphics[width=0.49\textwidth]{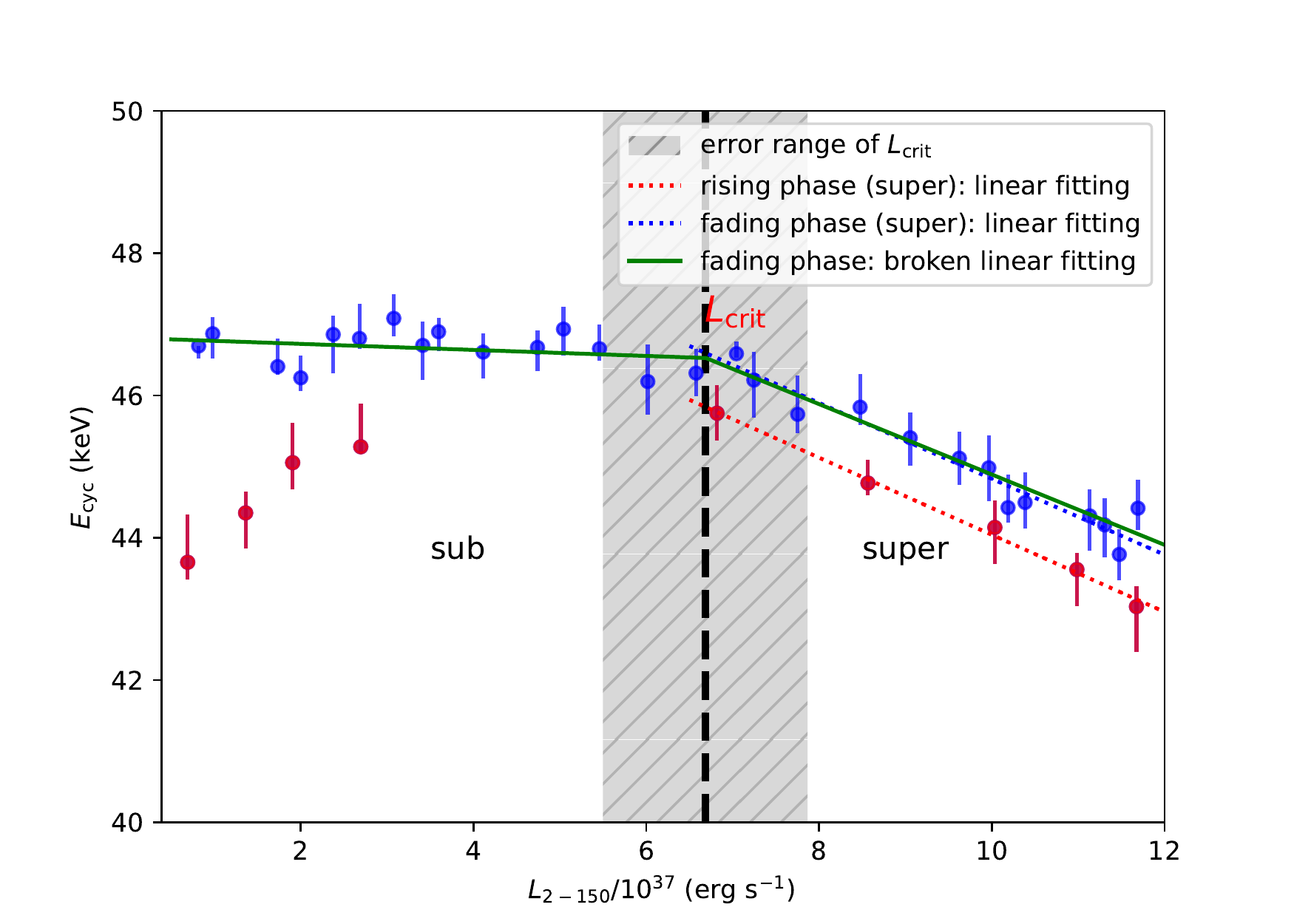}
    \centering\includegraphics[width=0.49\textwidth]{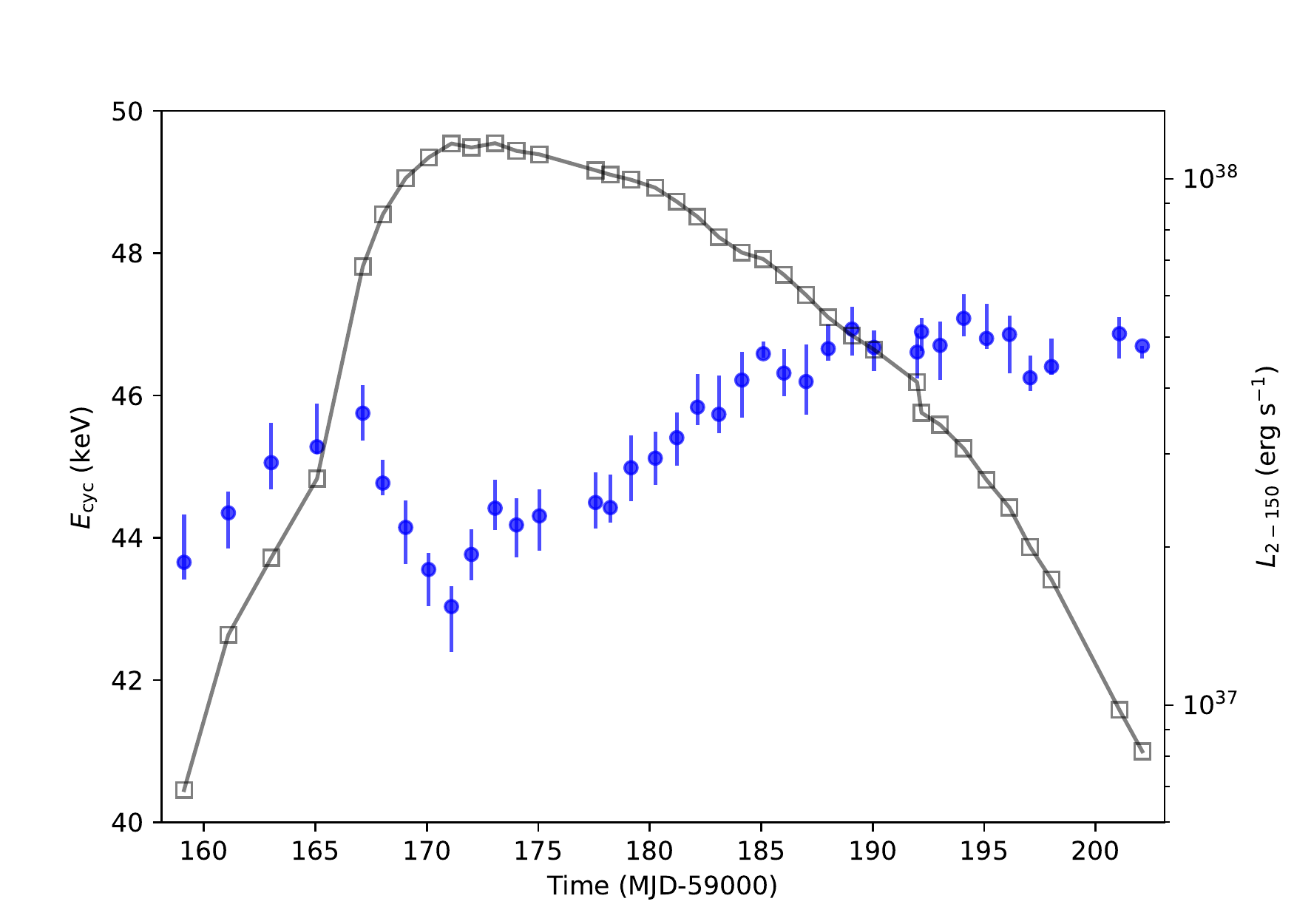}
    \caption{
    Left panel: 
    The evolution of the $E_{\rm cyc}$ with luminosity is shown. 
    The $E_{\rm cyc}$ in the rising and fading phase are shown in the red and blue colors, respectively. 
    The black dashed line shows the $L_{\rm break}$ with error range within grey shadow from a broken linear fitting, which consists to the critical luminosity $L_{\rm crit}$. 
    The red and blue dotted lines show the fit of the rising and fading trends of $E_{\rm cyc}$ during the super-critical state.
    Right panel: 
    The evolution of the fundamental cyclotron line $E_{\rm cyc}$ with time is shown with blue points. 
    The luminosity evolution with time is also shown with black points. 
    }
    \label{CRSF_Lum}
\end{figure}

To investigate the evolution of the spectral parameters with time and luminosity, the multi-component model described in the previous section was fitted to all Insight-HXMT spectra.
The left panel in Figure~\ref{CRSF_Lum} shows the variation of the fundamental line $E_{\rm cyc}$ with the $L_{x}$ (hereafter the luminosity is considered for the 2-150 keV energy range), where the $E_{\rm cyc}$ during rising phase and fading phase of the outburst are distinguished by blue and red points. 
The right panel in Figure~\ref{CRSF_Lum} shows the outline of the outburst with a resolution of one point per day.
The finer sampling of the fading phase allowed us to use a broken linear model to describe the relationship between $E_{\rm cyc}$ and $L_{x}$, which is shown with green line in the left panel of Figure~\ref{CRSF_Lum}.
The form of the broken linear model is:
\begin{equation}
E_{\rm cyc} = \left\{
\begin{array}{l}
k_1\times(\frac{L_{\rm x}}{10^{37} \rm erg\ s^{-1}})+(\frac{E_0}{\rm 1\ keV}),\ \ \ \rm if\ \frac{\emph{L}_{\rm x}}{10^{37} \rm erg\ s^{-1}} \leq \frac{\emph{L}_{\rm break}}{10^{37} \rm erg\ s^{-1}} \\
k_2\times(\frac{L_{\rm x}-L_{\rm break}}{10^{37} \rm erg\ s^{-1}})+k_1\times(\frac{L_{\rm break}}{10^{37} \rm erg\ s^{-1}})+(\frac{E_0}{\rm 1\ keV}),\ \ \ \rm if\ \frac{\emph{L}_{\rm x}}{10^{37} \rm erg\ s^{-1}} > \frac{\emph{L}_{\rm break}}{10^{37} \rm erg\ s^{-1}}
\end{array}
\right.
\end{equation}
where, $k_1$ and $k_2$ are the increasing/decreasing rate, and $L_{\rm break}$ is the break point.
From the fitting, we get $E_0=46.8\pm0.14 $ keV, $k_1=-0.04\pm0.04$, $k_2=0.49\pm0.05$ and $L_{\rm break}=6.7\pm0.4$.
The $E_{\rm cyc}$ and $L_{\rm x}$ follow a clear anti-correlation above $L_{\rm break}$, whereas $E_{\rm cyc}$ does not significantly change with $L_{x}$ below $L_{\rm break}$.
Considering that $E_{\rm cyc}$ strongly depends on the magnetic field and the conditions in the region near the neutron star surface where the accretion flow is decelerated, the change in the relationship between $E_{\rm cyc}$ and luminosity points to a change in the accretion regime and allows to tentatively associate the break luminosity with the critical luminosity corresponding to onset of an accretion column: $L_{\rm break}=L_{\rm crit}/10^{37}\ \rm erg\ s^{-1}=6.7\pm0.4$.

Note that there is an important difference in behavior of the line in the rising and the declining parts of the outburst, even for comparable luminosities.
In particular, there is a positive correlation of $E_{\rm cyc}$ and $L_{x}$ below $L_{\rm crit}$ during the rising phase whereas $E_{\rm cyc}$ does not change significantly with luminosity during the fading phase, remaining at $\approx $ 47 keV.
We note that both types of behaviour were not previously reported for 1A~0535+262, however, this is likely simply due to the lack of suitable observations.
Indeed, Insight-HXMT observations sampled the rising phase of the outburst in detail for the first time while past investigations of the line properties mostly focused on the declining part of the outburst. This might explain the non-detection of the correlation of the line energy with luminosity at low to moderate luminosities up to now.
On the other hand, luminosity range where the anti-correlation is observed was simply not observed previously with instruments capable of accurate measurement of the line energy since the current outburst of the source is the brightest in recent history.

It is interesting to note also that $E_{\rm cyc}$ is systematically lower during the rising phase than during the fading phase. 
Within the super-critical zone, the luminosity dependence of the line energy can be fitted by $E_{\rm cyc}=-0.54\pm0.03 \times (L_{x}/10^{37}\ \rm erg\ s^{-1})+49.5\pm0.3$ keV for the rising phase and $E_{\rm cyc}=-0.53\pm0.04 \times (L_{x}/10^{37}\ \rm erg\ s^{-1})+50.2\pm0.4$ keV for the fading phase.
Although there is uncertainty in the fit, a shift of 0.7 keV is shown in the left panel of Figure~\ref{CRSF_Lum}, while the slope of the correlation appears to be constant throughout the outburst. 
We note that this behavior appears to be similar to the line evolution reported by \cite{Cusumano2016,Doroshenko2017,Vybornov2018} for V~0332+53, however, with some important differences. 
In particular, in V~0332+53 the line energy was: 
a) lower in the declining phase than in the rising phase of the outburst (i.e. opposite to 1A~0535+262), and b) consistent with a linear decay in time which resulted in slightly different anti-correlation slopes \citep{Doroshenko2017}. 

We emphasize that for both sources the observed luminosity dependence of the cyclotron line energy is inconsistent with simple predictions of the available theoretical models. Indeed, regardless of the assumed model the line energy is expected to be related to height of the column, which in turn is expected to be proportional to accretion rate, so one expects same line energy for a fixed luminosity, i.e. no hysteresis as observed in V~0332+53 or 1A~0535+262. \cite{Doroshenko2017} argued that this discrepancy can be resolved if column height is also influenced by additional factors, for instance, magnetospheric radius defining the footprint area of the column and thus indirectly its height. \cite{Doroshenko2017, Vybornov2018} demonstrated through analysis of spin evolution and power spectra of the aperiodc variability that gradual recession of the accretion disk throughout the outburst (for a fixed luminosity) is indeed a plausible explanation for V~0332+53. To understand whether similar scenario can be realized also in 1A~0535+262, we considered, therefore, possible complex time and luminosity dependence of the line centroid energy also for this source.

In particular, to decouple time and luminosity dependence of the CRSF energy we adopted the same approach as \citep{Doroshenko2017,Vybornov2018}, i.e. introduced a linear time drift of the centroid energy in addition to the broken linear fit described above.
The right panel of the Figure~\ref{CRSF_Lum} shows the time dependence of the X-ray luminosity in $2-150$ keV (grey points) and the fundamental CRSF energy (blue points).  
As a result we found that additional linear increase of the line energy at a rate of $\sim0.071$ keV d$^{-1}$ indeed allows to describe the overall evolution of the line energy throughout the outburst. We emphasize again that none of the theoretical models predict any hysteresis effects, so our modeling is just an attempt to decouple the theoretically expected luminosity dependence of the line energy from other, yet unidentified effects. 
This allows to discuss them separately and better understand overall dependence of the line energy.

\begin{figure}
    \centering\includegraphics[scale=0.8]{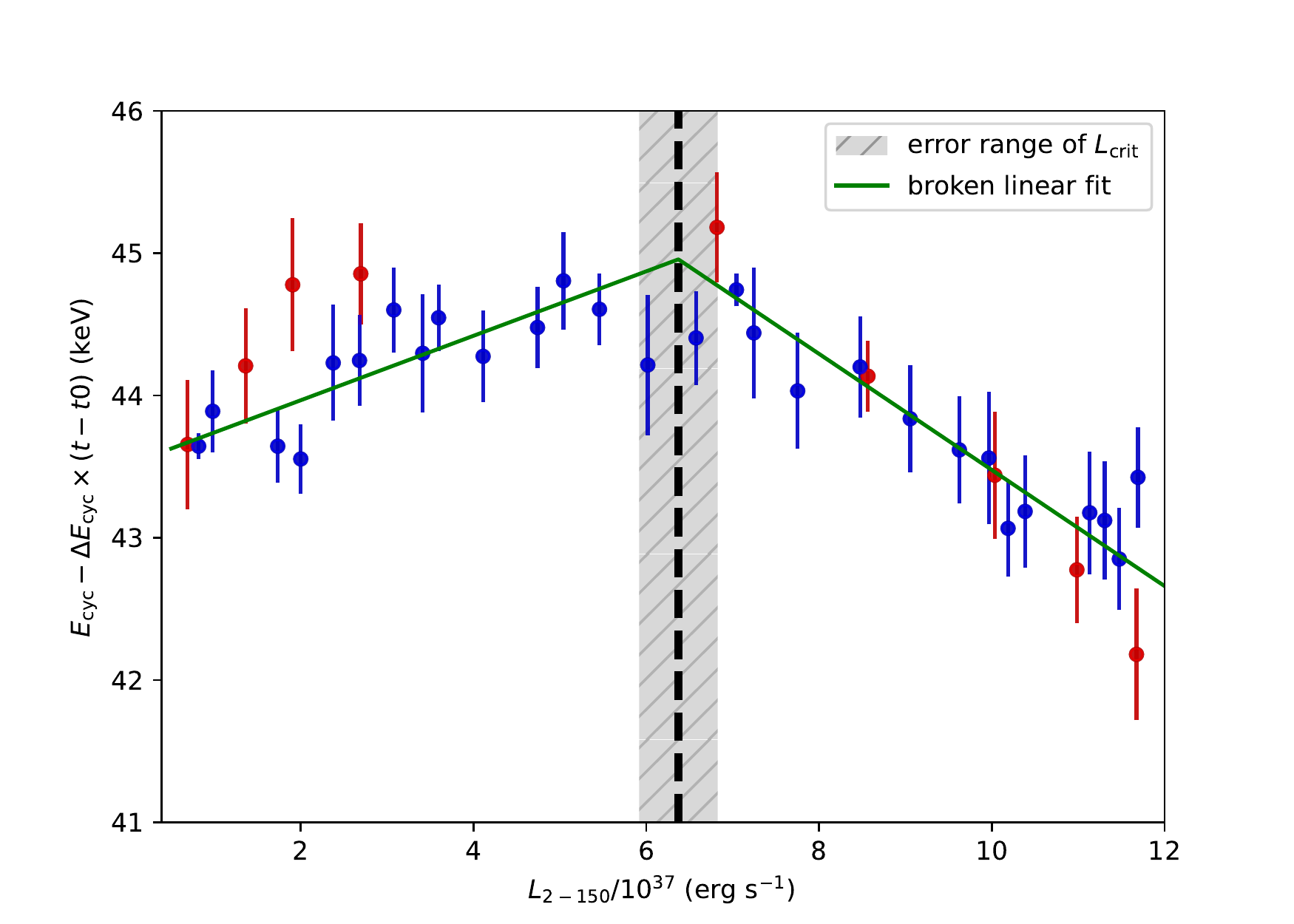}
    \caption{
    Correlation of the CRSF centroid energy on luminosity with the linear drift of $\Delta E_{\rm cyc}=0.071$ keV d$^{-1}$ taken into account.
    The $E_{\rm cyc}$ in the rising and fading phase are shown in the red and blue colors, respectively.
    The green line shows the broken linear fit.
    From the fit, the black dashed line shows the transition at $6.37\times 10^{38}$ erg s$^{-1}$. The grey area shows the error range of the transition luminosity. 
    }
    \label{L_t}
\end{figure}

First of all, we note that after subtracting this linear trend from the observed $E_{\rm cyc}-L_{\rm x}$ correlation the transition becomes more prominent as illustrated in Figure~\ref{L_t}. 
This allows better quantification of the observed properties of this correlation and more accurate measurement of the transition luminosity.
We estimate, i.e. slopes for the positive and negative correlation parts at $k_1=0.23\pm0.05$ and $k_2=-0.41\pm0.04$ with the break at the $L_{\rm break}=6.37\pm0.45$.
We note that observed luminosity of the break is consistent with theoretical expectations for onset of the accretion column (\citealp{Basko1976, Becker2012, Mushtukov2015MNRAS.447.1847M}) and thus interpret the observed transition as the accretion regime transition in 1A~0535+262.
Furthermore, the fact that, similarly to V~0332+53 additional time linear drift of the line energy on top of the expected luminosity depence is required, also deserves a discussion.

\subsection{Luminosity dependence of the broadband continuum spectrum}
\begin{figure}
    \centering\includegraphics[scale=0.8]{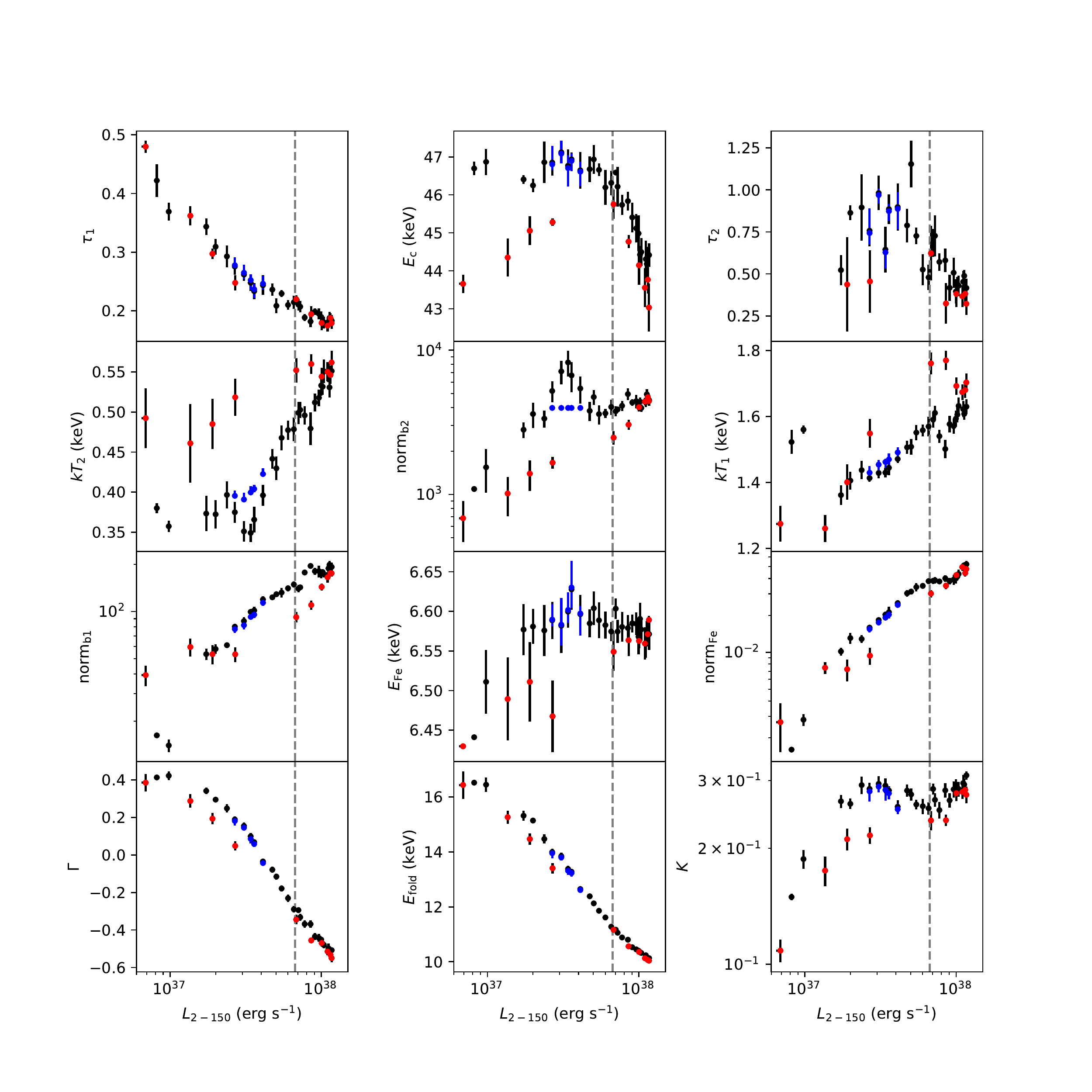}
    \caption{The evolution of the all parameters with the luminosity $L_{\rm x}$. The grey dashed line shows the critical luminosity $L_{\rm crit}$. The rising and fading phase during the outburst is distinguish with red and black points. The blue points show the corrected points by fixing $\rm norm_{\rm b, 2}$ at a reasonable value 4000}.
    \label{paras}
\end{figure}

In Figure~\ref{paras}, we plot the evolution of all spectral parameters as a function of luminosity during the outburst. 
The red and black points represent the rising and fading phase, respectively.
The grey dashed line denotes the critical luminosity at $6.7\times10^{38}$ erg s$^{-1}$.
As it is evident from the figure, the behavior in the rising and fading phases of the outburst is different for all parameters with the exception of $\tau_{1}$ of the fundamental line. 
Indeed, although the evolution with X-ray luminosity is smooth, the values of the parameters is significantly different at similar luminosities in the rising and fading phases.

The most prominent changes are seen in the black body components at higher luminosity.
For both black body components, the temperatures are systematically lower in the fading phase, and the normalisation parameters evolve in the opposite way.
From Figure~\ref{paras}, there is an unexpected hump of $\rm norm_{\rm b, 2}$ $\sim\ 4000-8000$ between $3-4 \times 10^{37}$ erg s$^{-1}$, and this hump does not match the change of the general trend. 
At this time, $kT_{2}$ evolves into a dip shape.
We note that this phenomenon may be caused by the degeneracy between temperature and normalization of the black body model rather than of physical origin.
By fixing $\rm norm_{\rm b,2}$ at 4000, we find that the evolution of $kT_{2}$ becomes smooth and continuous, and other parameters in the model did not change significantly (see blue points in Figure~\ref{paras}).
The spectrum at low energy, where the blackbody components dominate also display different trends with luminoisty (Figure~\ref{paras}).
The normalization of the blackbody is related to the size (radius) of the emitting area as norm=$R_{\rm BB}^{2}/d^{2}$, where $R_{\rm BB}$ is the radius if the emitting region in km and $d$ is the distance to the source in units of 10 kpc.
For lower temperature black body, the radius size can be limited to $R_{\rm BB}\sim5-13$ km, while for the higher temperature black body, it dominated at smaller radius size $R_{\rm BB}\sim1-3$ km.
The thermal emission radius varies $R_{\rm BB}\sim1-3$ km, which appears as a reasonable size for the polar cap.
The lower temperature black body could be associated with the top of the column cap (\citealp{Tao2019}).
The asymmetry in the evolution of these two black body components throughout the outburst may indicate that the accretion geometry is different in the rising and fading phase.

The spectrum tends to be harder (smaller $\Gamma$) and display smaller fold energies in the rising than in the fading phases.
We notice that the evolution trends of the spectral parameters in the black body and cutoff components do not remain at low luminosity, probably due to relatively poor statistics of the observational data.
The evolution of the harmonic CRSF at $\sim$ 100 keV (fixed at 2.3$\times E_{\rm cyc}$ for all observations) was not addressed in detail here because of the relatively high HE background at above 100 keV and hence larger systematic uncertainty in precisely measuring the 1-st harmonic CRSF parameter $\tau_{2}$.

\section{Discussion}
We have investigated the brightest outburst of 1A~0535+262 observed so far using the data from Insight-HXMT.
As a result, we found for the first time a clear anti-correlation between the cyclotron line energy and luminosity above $\sim6.7 \times 10^{37}$ erg s$^{-1}$ (see Figure~\ref{CRSF_Lum} and Figure~\ref{L_t}). 
For 1A~0535+262, this trend has not been found in the previous outbursts \citep{Staubert2019}, likely because the luminosity range where it occurs has not been sampled adequately by past observations.
This phenomenon is rare, and it has only been seen once before, i.e. in V~0332+53 (\citealp{Makishima1990}; \citealp{Tsygankov2010}; \citealp{Cusumano2016}; \citealp{Doroshenko2017} \citealp{Staubert2019}).
Thus, 1A~0535+262 is the second source that displays this behavior.
We emphasize that our investigation is the first to cover this luminosity range, which explains non-detection of the anti-correlation by previous studies.

The luminosity dependence of the observed line energy in both regimes has been already extensively discussed in the literature \citep{Tsygankov2007,Becker2012,Poutanen2013,Mushtukov2015MNRAS.447.1847M,Mushtukov2015MNRAS.454.2714M,Staubert2019}. 
Above the critical luminosity a higher accretion rate would result in larger radiation pressure and hence a higher accretion column above the magnetic pole and a smaller $E_{\rm cyc}$ no matter whether the line is observed directly from the column \citep{Tsygankov2007} or in reflection \citep{Poutanen2013}. 
In sub-critical regime the positive correlation between $E_{\rm cyc}$ and $L_{\rm x}$ can be explained if accreting matter is stopped via the Coulomb braking dominated process. 
In this case a collisionless shock first proposed by \cite{Langer1982} and \cite{Bykov2004} is expected to form at a height inversely proportional to the electron density in the emission region \citep{Shapiro1975}.
Therefore, height of the shock surface decreases with an accretion rate leading to a higher local magnetic field and correlation of the line energy with flux.
This model was successfully applied to interpret the $E_{\rm cyc}-L_{\rm x}$ correlations in Cep~X-4 (\citealp{Vybornov2017}) and GX~304-1 (\citealp{Rothschild2017}).
An alternative explanation for the observed CRSF energy evolution in the sub-critical regime is the Doppler effect caused by the radiation pressure which changes the velocity profile of the falling matter within the line-forming region (\citealp{Mushtukov2015MNRAS.454.2714M}).
When the radiation pressure increases with luminosity, the velocity near the NS surface decreases. 
Consequently, the lower electron velocity results in smaller redshift and higher CRSF energy. 
As a result, a positive correlation between the line centroid energy and the luminosity is expected in a regime with sub-critical luminosities.

As already mentioned in Section~3.2, however,  the observed evolution of line energy appears to be more complex than simple predictions outlined above. In particular, it is obvious from Figure~\ref{CRSF_Lum} that below $6.7 \times 10^{37}$ erg s$^{-1}$, the CRSF energy follows two different paths in the rising and fading phases.
In the rising phase, $E_{\rm cyc}$ is positively correlated with $L_{\rm x}$, while it is approximately constant in the fading phase.
We note that past studies of the source in this luminosity range were inconclusive, i.e. some had claimed a positive correlation \citep{Sartore2015}, while others reported the absence of any correlation (\citealp{Terada2006}; \citealp{Caballero2007}; \citealp{Caballero2013}). 
The observed behavior in the current outburst is consistent with these findings as the source was mostly observed during its outburst decline. The HXMT data confirms the absence of a correlation during the final stages of the outburst.

The overall observed line evolution throughout the outburst can be described as a combination of a time linear increase with a rate of $\Delta E_{\rm cyc}=0.071$ keV d$^{-1}$, and a broken linear dependence on luminosity shown in  Figure~\ref{L_t}. 
This allows us to estimate the transitional luminosity at $6.37\times10^{38}$\,erg\,s$^{-1}$.
We emphasize that this value is only valid under assumption above, i.e. that luminosity dependence of the line enrgy is modified by a time linear drift throughout the outburst. Nevertheless, the obtained value appears to be consistent with expected critical luminosity value associated with onset of an accretion column \citep{Basko1976, Becker2012, Mushtukov2015MNRAS.447.1847M}, which is estimated to be a broad range $2.0-6.7 \times10^{37}$ erg s$^{-1}$ (\citealp{Becker2012, Mushtukov2015MNRAS.447.1847M}). 
We conclude, therefore, that the observed break in luminosity dependence of the line energy is associated with the transition from sub-to super-critical accretion regime.

On the other hand, the origin of the time linear increase of the line energy remains unclear.
The asymmetric CRSF behavior was also reported for the 2005 outburst of V~0332+53 \citep{Cusumano2016,Doroshenko2017}, however, the CRSF energy in the rising part of the outburst was larger than in the fading phase in this case. 
\cite{Cusumano2016} interpreted the observed decrease  result as a rapid dissipation of external magnetic field of the neutron star due to diamagnetic screening effects caused by accreted matter. 
Alternatively, \cite{Doroshenko2017} argued that the energy drift of CRSF was related to the different emission geometries because of different magnetosphere sizes in rising and fading phase presumably associated with changes of internal structure of the accretion disk. 
We emphasize that unlike V~0332+53 the time linear drift in 1A~0535+262 implies increase rather than decrease of the line energy with time.
This implies that the scenario invoked by \cite{Cusumano2016} can definitively not be applied in this case and the observed changes in line energy must be associated with some change of the geometry or intrinsic conditions in the emission region rather any intrinsic field changes. 

To qualitatively understand what kind of changes might be responsible to account for the observed behavior of the line one could start by adopting scenario proposed by \cite{Doroshenko2017} for V~0332+53 reversing the rising and parts of the outburst. 
That is, to assume that for some reason the inner disk radius is larger and the polar cap footprint is smaller during the rising part of the outburst compared to those in the declining part. 
We emphasize that in this scenario it is not clear what could lead to different evolution of the inner disk radius in the two sources, and this discussion is out of scope of the current work. Therefore, below we only try to understand whether such change could in principle explain the observed line evolution or other explanation needs to be found.
With that said, one could anticipate (for a given luminosity) a taller accretion column in the rising phase, which is expected to result in a lower line energy no matter whether the line forms within the accretion column or in reflection off the neutron star surface as proposed by \cite{Poutanen2013}. 
We note that up to now no other scenarios have been proposed to explain anti-correlation of line energy with luminosity in super-critical state, so assumption that the polar cap indeed has a smaller footprint is, in fact, required in both scenarios to explain observed increase of the line energy in super-critical regime in the declining part of the outburst.

Considering that in the case of 1A~0535+262 the hysteresis in line energy is observed for both sub- and super-critical regimes, it is equally important, however, to consistently explain also the observed increase of the line energy in sub-critical part of the declining part of the outburst. 
As already mentioned, two scenarios to explain correlation of line energy with luminosity in the sub-critical regime have been proposed in the literature, i.e. change of the height of the collision-less shock \citep{Langer1982,Vybornov2017}, and Doppler shifts due to scattering in the accretion flow decelerated by radiative pressure \citep{Mushtukov2015MNRAS.454.2714M}. 
We note that in the latter model the lower footprint column in the rising phase, which can be characterized in terms of a larger $h/d$ (h is the height above the NS surface; d is the radius of the hotspot), is actually expected to lead to a lower observed line energy (see i.e. lower panel of Figure~\ref{specta_plt} in \citealt{Mushtukov2015MNRAS.454.2714M}). 
On the opposite, larger polar cap size required to explain higher line energy in super-critical state in the declining phase of the outburst, implies also lower $h/d$ and higher observed line energy also in sub-critical state. 
We conclude, therefore, that higher observed line energy (for a given luminosity) in the declining part of the outburst can be qualitatively explained by an increased footprint of the accretion flow. The larger normalization of the black body components during the declining phase would agree with this interpretation.

On the other hand, for collisionless shock model with higher observed line energy in the declining part corresponds to a stronger field in the line forming region and thus requires lower shock height resulting higher electron density in the scenario considered by \cite{Vybornov2018}. 
Recently, \cite{Kulsrud2020} proposed that the accretion matter can slip across the magnetic lines under a strong ideal Schwarzschild instability and hence increase the size of the polar cap. 
This would also imply lower column height of the collisionless shock, i.e. would also be qualitatively in agreement with observations. We note that increase of the polar cap size would also be consistent with the scenario by \cite{Mushtukov2015MNRAS.454.2714M}, i.e. it would be hard to disentangle between the two interpretations. 
On the other hand, leakage of matter through sides of the column proposed by \cite{Kulsrud2020} could provide physical explanation for the increase of the polar cap size in both cases.

We conclude, therefore, that the observed line energy is qualitatively consistent with the assumption of a larger inner disk radius in the rising phase of the outburst. We emphasize, however, that this conclusion is only one possibility and a detailed quantitative and self-consistent modeling of the observed line energy is required to really understand the behavior of the line in 1A~0535+262 and differences between this source and V~0332+53. 
Discussion of the latter is beyond the scope for the current outburst, however, we'd like to mention that the only obvious difference between the two pulsars is shorter spin period of V~0332+53. 
This implies that the latter source is likely much closer to co-rotation during an outburst, which could affect the interaction of the accretion flow with the magnetosphere and potentially alter magnetosphere size and footprint of the accretion column. 

\section{Conclusions}
The high cadence and counting statistics provided by observations of a giant outburst of 1A~0535+262 with Insight-HXMT in a broad energy range revealed for the first time a series of new findings in context of variability of the properties of the cyclotron line observed in this source with luminosity and time.
First, we were able to detect for the first time a clear anti-correlation of the observed line energy with luminosity above $\sim6.7\times10^{37}$\,erg\,s$^{-1}$. 
Furthermore, also for the first time, a clear correlation of the line energy with flux was observed below this luminosity in the rising part of the outburst. 

Overall, the observed line energy was found to exhibit complex behaviour characterized by apparent hysteresis of line parameters between rising and declining parts of the outburst. This hysteresis can be accounted for by a time-linear increase of the line energy at a rate of $0.071$ keV d$^{-1}$ on-top of a combination of a correlation with flux below critical luminosity of $\sim6.7\times10^{37}$\,erg\,s$^{-1}$ and anti-correlation above this level. This behaviour is similar to that previously reported for another Be-transient V~0332+53 \citep{Doroshenko2017,Vybornov2018}, although in the latter case a time linear decay rather than increase was observed. The origin of this discrepancy is unclear and subject for further investigations, however, already now our results clearly show that the observed variability of cyclotron lines in X-ray pulsars is clearly more complex than was foreseen by theoretical models. 

\acknowledgments
This work made use of data from the Insight-HXMT mission, a project funded by China National Space Administration (CNSA) and the Chinese Academy of Sciences (CAS).
This work is supported by the National Key R\&D Program of China (2016YFA0400800) and the National Natural Science Foundation of China under grants U1838201, 11473027, U1838202, 11733009, U1838104, U1938101, U2038101, U1938103 and Guangdong Major Project of Basic and Applied Basic Research (Grant No. 2019B030302001).

\bibliography{ref}

\begin{thebibliography}{}
\expandafter\ifx\csname natexlab\endcsname\relax\def\natexlab#1{#1}\fi
\providecommand{\url}[1]{\href{#1}{#1}}
\providecommand{\dodoi}[1]{doi:~\href{http://doi.org/#1}{\nolinkurl{#1}}}
\providecommand{\doeprint}[1]{\href{http://ascl.net/#1}{\nolinkurl{http://ascl.net/#1}}}
\providecommand{\doarXiv}[1]{\href{https://arxiv.org/abs/#1}{\nolinkurl{https://arxiv.org/abs/#1}}}

\bibitem[{{Arnaud}(1996)}]{1996ASPC..101...17A}
{Arnaud}, K.~A. 1996, in Astronomical Society of the Pacific Conference Series,
  Vol. 101, Astronomical Data Analysis Software and Systems V, ed. G.~H.
  {Jacoby} \& J.~{Barnes}, 17

\bibitem[{{Bailer-Jones} {et~al.}(2018){Bailer-Jones}, {Rybizki}, {Fouesneau},
  {Mantelet}, \& {Andrae}}]{Bailer-Jones2018}
{Bailer-Jones}, C.~A.~L., {Rybizki}, J., {Fouesneau}, M., {Mantelet}, G., \&
  {Andrae}, R. 2018, \aj, 156, 58, \dodoi{10.3847/1538-3881/aacb21}

\bibitem[{{Basko} \& {Sunyaev}(1976)}]{Basko1976}
{Basko}, M.~M., \& {Sunyaev}, R.~A. 1976, \mnras, 175, 395,
  \dodoi{10.1093/mnras/175.2.395}

\bibitem[{{Becker} {et~al.}(2012){Becker}, {Klochkov}, {Sch{\"o}nherr},
  {Nishimura}, {Ferrigno}, {Caballero}, {Kretschmar}, {Wolff}, {Wilms}, \&
  {Staubert}}]{Becker2012}
{Becker}, P.~A., {Klochkov}, D., {Sch{\"o}nherr}, G., {et~al.} 2012, \aap, 544,
  A123, \dodoi{10.1051/0004-6361/201219065}

\bibitem[{{Bykov} \& {Krasilshchikov}(2004)}]{Bykov2004}
{Bykov}, A.~M., \& {Krasilshchikov}, A.~M. 2004, Astronomy Letters, 30, 309,
  \dodoi{10.1134/1.1738153}

\bibitem[{{Caballero} {et~al.}(2007){Caballero}, {Kretschmar}, {Santangelo},
  {Staubert}, {Klochkov}, {Camero}, {Ferrigno}, {Finger}, {Kreykenbohm},
  {McBride}, {Pottschmidt}, {Rothschild}, {Sch{\"o}nherr}, {Segreto}, {Suchy},
  {Wilms}, \& {Wilson}}]{Caballero2007}
{Caballero}, I., {Kretschmar}, P., {Santangelo}, A., {et~al.} 2007, \aap, 465,
  L21, \dodoi{10.1051/0004-6361:20067032}

\bibitem[{{Caballero} {et~al.}(2008){Caballero}, {Santangelo}, {Kretschmar},
  {Staubert}, {Postnov}, {Klochkov}, {Camero-Arranz}, {Finger}, {Kreykenbohm},
  {Pottschmidt}, {Rothschild}, {Suchy}, {Wilms}, \& {Wilson}}]{Caballero2008}
{Caballero}, I., {Santangelo}, A., {Kretschmar}, P., {et~al.} 2008, \aap, 480,
  L17, \dodoi{10.1051/0004-6361:20079310}

\bibitem[{{Caballero} {et~al.}(2013){Caballero}, {Pottschmidt}, {Marcu},
  {Barragan}, {Ferrigno}, {Klochkov}, {Zurita Heras}, {Suchy}, {Wilms},
  {Kretschmar}, {Santangelo}, {Kreykenbohm}, {F{\"u}rst}, {Rothschild},
  {Staubert}, {Finger}, {Camero-Arranz}, {Makishima}, {Enoto}, {Iwakiri}, \&
  {Terada}}]{Caballero2013}
{Caballero}, I., {Pottschmidt}, K., {Marcu}, D.~M., {et~al.} 2013, \apjl, 764,
  L23, \dodoi{10.1088/2041-8205/764/2/L23}

\bibitem[{{Camero-Arranz} {et~al.}(2012){Camero-Arranz}, {Finger},
  {Wilson-Hodge}, {Jenke}, {Steele}, {Coe}, {Gutierrez-Soto}, {Kretschmar},
  {Caballero}, {Yan}, {Rodr{\'\i}guez}, {Suso}, {Case}, {Cherry}, {Guiriec}, \&
  {McBride}}]{Camero-Arranz2012}
{Camero-Arranz}, A., {Finger}, M.~H., {Wilson-Hodge}, C.~A., {et~al.} 2012,
  \apj, 754, 20, \dodoi{10.1088/0004-637X/754/1/20}

\bibitem[{{Cao} {et~al.}(2020){Cao}, {Jiang}, {Meng}, {Zhang}, {Luo}, {Yang},
  {Zhang}, {Gu}, {Sun}, {Liu}, {Yang}, {Li}, {Tan}, {Liu}, {Du}, {Lu}, {Xu},
  {Guan}, {Zhang}, {Wang}, {Li}, {Zhang}, {Wen}, {Qu}, {Song}, {Li}, {Ge},
  {Zhou}, {Xiong}, {Zhang}, {Zhang}, {Cheng}, {Zhang}, {Li}, {Liang}, {Gao},
  {Yang}, {Liu}, {Liu}, {Yang}, \& {Zhang}}]{2020SCPMA..63x9504C}
{Cao}, X., {Jiang}, W., {Meng}, B., {et~al.} 2020, Science China Physics,
  Mechanics, and Astronomy, 63, 249504, \dodoi{10.1007/s11433-019-1506-1}

\bibitem[{{Chen} {et~al.}(2020){Chen}, {Cui}, {Li}, {Wang}, {Xu}, {Lu}, {Wang},
  {Chen}, {Han}, {Hu}, {Zhang}, {Huo}, {Yang}, {Li}, {Lu}, {Zhang}, {Li},
  {Zhang}, {Xiong}, {Zhang}, {Xue}, {Zhao}, {Zhu}, {Zhu}, {Liu}, {Yang}, \&
  {Zhang}}]{2020SCPMA..63x9505C}
{Chen}, Y., {Cui}, W., {Li}, W., {et~al.} 2020, Science China Physics,
  Mechanics, and Astronomy, 63, 249505, \dodoi{10.1007/s11433-019-1469-5}

\bibitem[{{Cusumano} {et~al.}(2016){Cusumano}, {La Parola}, {D'A{\`\i}},
  {Segreto}, {Tagliaferri}, {Barthelmy}, \& {Gehrels}}]{Cusumano2016}
{Cusumano}, G., {La Parola}, V., {D'A{\`\i}}, A., {et~al.} 2016, \mnras, 460,
  L99, \dodoi{10.1093/mnrasl/slw084}

\bibitem[{{Doroshenko} {et~al.}(2017){Doroshenko}, {Tsygankov}, {Mushtukov},
  {Lutovinov}, {Santangelo}, {Suleimanov}, \& {Poutanen}}]{Doroshenko2017}
{Doroshenko}, V., {Tsygankov}, S.~S., {Mushtukov}, A.~A., {et~al.} 2017,
  \mnras, 466, 2143, \dodoi{10.1093/mnras/stw3236}

\bibitem[{{Finger} {et~al.}(1996){Finger}, {Wilson}, \& {Harmon}}]{Finger1996}
{Finger}, M.~H., {Wilson}, R.~B., \& {Harmon}, B.~A. 1996, \apj, 459, 288,
  \dodoi{10.1086/176892}

\bibitem[{{F{\"u}rst} {et~al.}(2014){F{\"u}rst}, {Pottschmidt}, {Wilms},
  {Tomsick}, {Bachetti}, {Boggs}, {Christensen}, {Craig}, {Grefenstette},
  {Hailey}, {Harrison}, {Madsen}, {Miller}, {Stern}, {Walton}, \&
  {Zhang}}]{Furst2014}
{F{\"u}rst}, F., {Pottschmidt}, K., {Wilms}, J., {et~al.} 2014, \apj, 780, 133,
  \dodoi{10.1088/0004-637X/780/2/133}

\bibitem[{{F{\"u}rst} {et~al.}(2015){F{\"u}rst}, {Pottschmidt}, {Miyasaka},
  {Bhalerao}, {Bachetti}, {Boggs}, {Christensen}, {Craig}, {Grinberg},
  {Hailey}, {Harrison}, {Kennea}, {Rahoui}, {Stern}, {Tendulkar}, {Tomsick},
  {Walton}, {Wilms}, \& {Zhang}}]{Furst2015}
{F{\"u}rst}, F., {Pottschmidt}, K., {Miyasaka}, H., {et~al.} 2015, \apjl, 806,
  L24, \dodoi{10.1088/2041-8205/806/2/L24}

\bibitem[{{Grove} {et~al.}(1995){Grove}, {Strickman}, {Johnson}, {Kurfess},
  {Kinzer}, {Starr}, {Jung}, {Kendziorra}, {Kretschmar}, {Maisack}, \&
  {Staubert}}]{Grove1995}
{Grove}, J.~E., {Strickman}, M.~S., {Johnson}, W.~N., {et~al.} 1995, \apjl,
  438, L25, \dodoi{10.1086/187706}

\bibitem[{{Guo} {et~al.}(2020){Guo}, {Liao}, {Zhang}, {Zhang}, {Tan}, {Song},
  {Lu}, {Cao}, {Chang}, {Chen}, {Du}, {Ge}, {Gu}, {Jiang}, {Li}, {Li}, {Li},
  {Liu}, {Liu}, {Lu}, {Luo}, {Meng}, {Sun}, {Yang}, {Yang}, {You}, {Zhang},
  {Zhao}, \& {Zhang}}]{2020arXiv200306260G}
{Guo}, C.-C., {Liao}, J.-Y., {Zhang}, S., {et~al.} 2020, arXiv e-prints,
  arXiv:2003.06260.
\newblock \doarXiv{2003.06260}

\bibitem[{{Iyer} {et~al.}(2015){Iyer}, {Mukherjee}, {Dewangan}, {Bhattacharya},
  \& {Seetha}}]{Iyer2015}
{Iyer}, N., {Mukherjee}, D., {Dewangan}, G.~C., {Bhattacharya}, D., \&
  {Seetha}, S. 2015, \mnras, 454, 741, \dodoi{10.1093/mnras/stv1942}

\bibitem[{{Kendziorra} {et~al.}(1994){Kendziorra}, {Kretschmar}, {Pan}, {Kunz},
  {Maisack}, {Staubert}, {Pietsch}, {Truemper}, {Efremov}, \&
  {Sunyaev}}]{Kendziorra1994}
{Kendziorra}, E., {Kretschmar}, P., {Pan}, H.~C., {et~al.} 1994, \aap, 291, L31

\bibitem[{{Klochkov} {et~al.}(2011){Klochkov}, {Staubert}, {Santangelo},
  {Rothschild}, \& {Ferrigno}}]{Klochkov2011}
{Klochkov}, D., {Staubert}, R., {Santangelo}, A., {Rothschild}, R.~E., \&
  {Ferrigno}, C. 2011, \aap, 532, A126, \dodoi{10.1051/0004-6361/201116800}

\bibitem[{{Klochkov} {et~al.}(2012){Klochkov}, {Doroshenko}, {Santangelo},
  {Staubert}, {Ferrigno}, {Kretschmar}, {Caballero}, {Wilms}, {Kreykenbohm},
  {Pottschmidt}, {Rothschild}, {Wilson-Hodge}, \&
  {P{\"u}hlhofer}}]{Klochkov2012}
{Klochkov}, D., {Doroshenko}, V., {Santangelo}, A., {et~al.} 2012, \aap, 542,
  L28, \dodoi{10.1051/0004-6361/201219385}

\bibitem[{{Kulsrud} \& {Sunyaev}(2020)}]{Kulsrud2020}
{Kulsrud}, R.~M., \& {Sunyaev}, R. 2020, Journal of Plasma Physics, 86,
  905860602, \dodoi{10.1017/S0022377820001026}

\bibitem[{{La Parola} {et~al.}(2016){La Parola}, {Cusumano}, {Segreto}, \&
  {D'A{\`\i}}}]{Parola2016}
{La Parola}, V., {Cusumano}, G., {Segreto}, A., \& {D'A{\`\i}}, A. 2016,
  \mnras, 463, 185, \dodoi{10.1093/mnras/stw1915}

\bibitem[{{Langer} \& {Rappaport}(1982)}]{Langer1982}
{Langer}, S.~H., \& {Rappaport}, S. 1982, \apj, 257, 733,
  \dodoi{10.1086/160028}

\bibitem[{{Liao} {et~al.}(2020{\natexlab{a}}){Liao}, {Zhang}, {Chen}, {Zhang},
  {Jin}, {Chang}, {Chen}, {Ge}, {Guo}, {Li}, {Li}, {Lu}, {Lu}, {Nie}, {Song},
  {Yang}, {You}, {Zhao}, \& {Zhang}}]{2020arXiv200401432L}
{Liao}, J.-Y., {Zhang}, S., {Chen}, Y., {et~al.} 2020{\natexlab{a}}, arXiv
  e-prints, arXiv:2004.01432.
\newblock \doarXiv{2004.01432}

\bibitem[{{Liao} {et~al.}(2020{\natexlab{b}}){Liao}, {Zhang}, {Lu}, {Zhang},
  {Li}, {Chang}, {Chen}, {Ge}, {Guo}, {Huang}, {Jin}, {Li}, {Li}, {Li}, {Liu},
  {Lu}, {Nie}, {Song}, {Wang}, {You}, {Zhang}, {Zhao}, \&
  {Zhang}}]{2020arXiv200501661L}
{Liao}, J.-Y., {Zhang}, S., {Lu}, X.-F., {et~al.} 2020{\natexlab{b}}, arXiv
  e-prints, arXiv:2005.01661.
\newblock \doarXiv{2005.01661}

\bibitem[{{Liu} {et~al.}(2020){Liu}, {Zhang}, {Li}, {Lu}, {Chang}, {Li},
  {Zhang}, {Jin}, {Yu}, {Zhang}, {Fu}, {Chen}, {Ji}, {Xu}, {Deng}, {Shang},
  {Liu}, {Lu}, {Zhang}, {Dong}, {Li}, {Wu}, {Li}, {Wang}, {Wu}, {Zhang},
  {Zhang}, {Xiong}, {Liu}, {Zhang}, {Liu}, {Yang}, \&
  {Zhang}}]{2020SCPMA..63x9503L}
{Liu}, C., {Zhang}, Y., {Li}, X., {et~al.} 2020, Science China Physics,
  Mechanics, and Astronomy, 63, 249503, \dodoi{10.1007/s11433-019-1486-x}

\bibitem[{{Makishima} {et~al.}(1990){Makishima}, {Mihara}, {Ishida}, {Ohashi},
  {Sakao}, {Tashiro}, {Tsuru}, {Kii}, {Makino}, {Murakami}, {Nagase}, {Tanaka},
  {Kunieda}, {Tawara}, {Kitamoto}, {Miyamoto}, {Yoshida}, \&
  {Turner}}]{Makishima1990}
{Makishima}, K., {Mihara}, T., {Ishida}, M., {et~al.} 1990, \apjl, 365, L59,
  \dodoi{10.1086/185888}

\bibitem[{{M{\"u}ller} {et~al.}(2013){M{\"u}ller}, {Ferrigno}, {K{\"u}hnel},
  {Sch{\"o}nherr}, {Becker}, {Wolff}, {Hertel}, {Schwarm}, {Grinberg}, {Obst},
  {Caballero}, {Pottschmidt}, {F{\"u}rst}, {Kreykenbohm}, {Rothschild},
  {Hemphill}, {N{\'u}{\~n}ez}, {Torrej{\'o}n}, {Klochkov}, {Staubert}, \&
  {Wilms}}]{Mueller2013}
{M{\"u}ller}, S., {Ferrigno}, C., {K{\"u}hnel}, M., {et~al.} 2013, \aap, 551,
  A6, \dodoi{10.1051/0004-6361/201220359}

\bibitem[{{Mushtukov} {et~al.}(2015{\natexlab{a}}){Mushtukov}, {Suleimanov},
  {Tsygankov}, \& {Poutanen}}]{Mushtukov2015MNRAS.447.1847M}
{Mushtukov}, A.~A., {Suleimanov}, V.~F., {Tsygankov}, S.~S., \& {Poutanen}, J.
  2015{\natexlab{a}}, \mnras, 447, 1847, \dodoi{10.1093/mnras/stu2484}

\bibitem[{{Mushtukov} {et~al.}(2015{\natexlab{b}}){Mushtukov}, {Tsygankov},
  {Serber}, {Suleimanov}, \& {Poutanen}}]{Mushtukov2015MNRAS.454.2714M}
{Mushtukov}, A.~A., {Tsygankov}, S.~S., {Serber}, A.~V., {Suleimanov}, V.~F.,
  \& {Poutanen}, J. 2015{\natexlab{b}}, \mnras, 454, 2714,
  \dodoi{10.1093/mnras/stv2182}

\bibitem[{{Nakajima} {et~al.}(2006){Nakajima}, {Mihara}, {Makishima}, \&
  {Niko}}]{Nakajima2006}
{Nakajima}, M., {Mihara}, T., {Makishima}, K., \& {Niko}, H. 2006, \apj, 646,
  1125, \dodoi{10.1086/502638}

\bibitem[{{Postnov} {et~al.}(2008){Postnov}, {Staubert}, {Santangelo},
  {Klochkov}, {Kretschmar}, \& {Caballero}}]{Postnov2008}
{Postnov}, K., {Staubert}, R., {Santangelo}, A., {et~al.} 2008, \aap, 480, L21,
  \dodoi{10.1051/0004-6361:20079277}

\bibitem[{{Poutanen} {et~al.}(2013){Poutanen}, {Mushtukov}, {Suleimanov},
  {Tsygankov}, {Nagirner}, {Doroshenko}, \& {Lutovinov}}]{Poutanen2013}
{Poutanen}, J., {Mushtukov}, A.~A., {Suleimanov}, V.~F., {et~al.} 2013, \apj,
  777, 115, \dodoi{10.1088/0004-637X/777/2/115}

\bibitem[{{Reig} \& {Nespoli}(2013)}]{Reig2013}
{Reig}, P., \& {Nespoli}, E. 2013, \aap, 551, A1,
  \dodoi{10.1051/0004-6361/201219806}

\bibitem[{{Rosenberg} {et~al.}(1975){Rosenberg}, {Eyles}, {Skinner}, \&
  {Willmore}}]{Rosenberg1975}
{Rosenberg}, F.~D., {Eyles}, C.~J., {Skinner}, G.~K., \& {Willmore}, A.~P.
  1975, \nat, 256, 628, \dodoi{10.1038/256628a0}

\bibitem[{{Rothschild} {et~al.}(2017){Rothschild}, {K{\"u}hnel}, {Pottschmidt},
  {Hemphill}, {Postnov}, {Gornostaev}, {Shakura}, {F{\"u}rst}, {Wilms},
  {Staubert}, \& {Klochkov}}]{Rothschild2017}
{Rothschild}, R.~E., {K{\"u}hnel}, M., {Pottschmidt}, K., {et~al.} 2017,
  \mnras, 466, 2752, \dodoi{10.1093/mnras/stw3222}

\bibitem[{{Sartore} {et~al.}(2015){Sartore}, {Jourdain}, \&
  {Roques}}]{Sartore2015}
{Sartore}, N., {Jourdain}, E., \& {Roques}, J.~P. 2015, \apj, 806, 193,
  \dodoi{10.1088/0004-637X/806/2/193}

\bibitem[{{Shapiro} \& {Salpeter}(1975)}]{Shapiro1975}
{Shapiro}, S.~L., \& {Salpeter}, E.~E. 1975, \apj, 198, 671,
  \dodoi{10.1086/153645}

\bibitem[{{Staubert} {et~al.}(2007){Staubert}, {Shakura}, {Postnov}, {Wilms},
  {Rothschild}, {Coburn}, {Rodina}, \& {Klochkov}}]{Staubert2007}
{Staubert}, R., {Shakura}, N.~I., {Postnov}, K., {et~al.} 2007, \aap, 465, L25,
  \dodoi{10.1051/0004-6361:20077098}

\bibitem[{{Staubert} {et~al.}(2019){Staubert}, {Tr{\"u}mper}, {Kendziorra},
  {Klochkov}, {Postnov}, {Kretschmar}, {Pottschmidt}, {Haberl}, {Rothschild},
  {Santangelo}, {Wilms}, {Kreykenbohm}, \& {F{\"u}rst}}]{Staubert2019}
{Staubert}, R., {Tr{\"u}mper}, J., {Kendziorra}, E., {et~al.} 2019, \aap, 622,
  A61, \dodoi{10.1051/0004-6361/201834479}

\bibitem[{{Steele} {et~al.}(1998){Steele}, {Negueruela}, {Coe}, \&
  {Roche}}]{Steele1998}
{Steele}, I.~A., {Negueruela}, I., {Coe}, M.~J., \& {Roche}, P. 1998, \mnras,
  297, L5, \dodoi{10.1046/j.1365-8711.1998.01593.x}

\bibitem[{{Tao} {et~al.}(2019){Tao}, {Feng}, {Zhang}, {Bu}, {Zhang}, {Qu}, \&
  {Zhang}}]{Tao2019}
{Tao}, L., {Feng}, H., {Zhang}, S., {et~al.} 2019, \apj, 873, 19,
  \dodoi{10.3847/1538-4357/ab0211}

\bibitem[{{Terada} {et~al.}(2006){Terada}, {Mihara}, {Nakajima}, {Suzuki},
  {Isobe}, {Makishima}, {Takahashi}, {Enoto}, {Kokubun}, {Kitaguchi}, {Naik},
  {Dotani}, {Nagase}, {Tanaka}, {Watanabe}, {Kitamoto}, {Sudoh}, {Yoshida},
  {Nakagawa}, {Sugita}, {Kohmura}, {Kotani}, {Yonetoku}, {Angelini}, {Cottam},
  {Mukai}, {Kelley}, {Soong}, {Bautz}, {Kissel}, \& {Doty}}]{Terada2006}
{Terada}, Y., {Mihara}, T., {Nakajima}, M., {et~al.} 2006, \apjl, 648, L139,
  \dodoi{10.1086/508018}

\bibitem[{{Tsygankov} {et~al.}(2007){Tsygankov}, {Lutovinov}, {Churazov}, \&
  {Sunyaev}}]{Tsygankov2007}
{Tsygankov}, S.~S., {Lutovinov}, A.~A., {Churazov}, E.~M., \& {Sunyaev}, R.~A.
  2007, Astronomy Letters, 33, 368, \dodoi{10.1134/S1063773707060023}

\bibitem[{{Tsygankov} {et~al.}(2010){Tsygankov}, {Lutovinov}, \&
  {Serber}}]{Tsygankov2010}
{Tsygankov}, S.~S., {Lutovinov}, A.~A., \& {Serber}, A.~V. 2010, \mnras, 401,
  1628, \dodoi{10.1111/j.1365-2966.2009.15791.x}

\bibitem[{{Vybornov} {et~al.}(2018){Vybornov}, {Doroshenko}, {Staubert}, \&
  {Santangelo}}]{Vybornov2018}
{Vybornov}, V., {Doroshenko}, V., {Staubert}, R., \& {Santangelo}, A. 2018,
  \aap, 610, A88, \dodoi{10.1051/0004-6361/201731750}

\bibitem[{{Vybornov} {et~al.}(2017){Vybornov}, {Klochkov}, {Gornostaev},
  {Postnov}, {Sokolova-Lapa}, {Staubert}, {Pottschmidt}, \&
  {Santangelo}}]{Vybornov2017}
{Vybornov}, V., {Klochkov}, D., {Gornostaev}, M., {et~al.} 2017, \aap, 601,
  A126, \dodoi{10.1051/0004-6361/201630275}

\bibitem[{{Wilms} {et~al.}(2000){Wilms}, {Allen}, \& {McCray}}]{Wilms2000}
{Wilms}, J., {Allen}, A., \& {McCray}, R. 2000, \apj, 542, 914,
  \dodoi{10.1086/317016}

\bibitem[{{Zhang} {et~al.}(2014){Zhang}, {Lu}, {Zhang}, \&
  {Li}}]{2014SPIE.9144E..21Z}
{Zhang}, S., {Lu}, F.~J., {Zhang}, S.~N., \& {Li}, T.~P. 2014, in Society of
  Photo-Optical Instrumentation Engineers (SPIE) Conference Series, Vol. 9144,
  Space Telescopes and Instrumentation 2014: Ultraviolet to Gamma Ray, 914421,
  \dodoi{10.1117/12.2054144}

\bibitem[{{Zhang} {et~al.}(2020){Zhang}, {Li}, {Lu}, {Song}, {Xu}, {Liu},
  {Chen}, {Cao}, {Bu}, {Chang}, {Chen}, {Chen}, {Chen}, {Chen}, {Chen}, {Cui},
  {Cui}, {Deng}, {Dong}, {Du}, {Fu}, {Gao}, {Gao}, {Gao}, {Ge}, {Gu}, {Guan},
  {Gungor}, {Guo}, {Han}, {Hu}, {Huang}, {Huo}, {Jia}, {Jiang}, {Jiang}, {Jin},
  {Jin}, {Li}, {Li}, {Li}, {Li}, {Li}, {Li}, {Li}, {Li}, {Li}, {Li}, {Li},
  {Liang}, {Liao}, {Liu}, {Liu}, {Liu}, {Liu}, {Liu}, {Liu}, {Lu}, {Lu}, {Luo},
  {Ma}, {Meng}, {Nang}, {Nie}, {Ou}, {Qu}, {Sai}, {Shang}, {Shen}, {Sun},
  {Tan}, {Tao}, {Tuo}, {Wang}, {Wang}, {Wang}, {Wang}, {Wang}, {Wang}, {Wang},
  {Wen}, {Wu}, {Wu}, {Wu}, {Xiao}, {Xiong}, {Yan}, {Yang}, {Yang}, {Yang},
  {Yi}, {Yuan}, {Zhang}, {Zhang}, {Zhang}, {Zhang}, {Zhang}, {Zhang}, {Zhang},
  {Zhang}, {Zhang}, {Zhang}, {Zhang}, {Zhang}, {Zhang}, {Zhang}, {Zhang},
  {Zhang}, {Zhang}, {Zhang}, {Zhang}, {Zhang}, {Zhao}, {Zhao}, {Zheng}, {Zhou},
  {Zhu}, {Zhu}, {Zhuang}, \& {Insight-HXMT team}}]{2020SCPMA..63x9502Z}
{Zhang}, S.-N., {Li}, T., {Lu}, F., {et~al.} 2020, Science China Physics,
  Mechanics, and Astronomy, 63, 249502, \dodoi{10.1007/s11433-019-1432-6}

\end{thebibliography}
\bibliographystyle{aasjournal}

\end{document}